\documentclass[a4paper,11pt]{article}

\usepackage[utf8]{inputenc}
\usepackage[T1]{fontenc}
\usepackage[english]{babel}
\usepackage{jcappub} \makeatletter\gdef\@fpheader{}\makeatletter
\usepackage{ragged2e}
\usepackage{microtype}
\usepackage{lineno}
\usepackage[nottoc]{tocbibind}
\usepackage{amsfonts}
\usepackage{amsthm}
\usepackage{mathtools}
\usepackage{bm}
\usepackage{enumitem}
\usepackage{slashed}
\usepackage{float}
\usepackage{tikz}
\usepackage{orcidlink}
\usepackage{subcaption}
\usepackage[
  font=small,
  labelfont=bf,
  labelsep=period,
  justification=justified,
  singlelinecheck=false
]{caption}
\usepackage{adjustbox}
\usepackage{booktabs}
\usepackage{tabularx}
\usepackage{placeins}

\hypersetup{pdftitle={Strong First-Order Electroweak Phase Transition and Gravitational Waves in a Z4 Fermion--Scalar Dark Matter Model},
            pdfauthor={J. P. Carvalho-Corrêa, J. P. Cunha-Melo, I. M. Pereira, B. L. Sánchez-Vega, A. C. D. Viglioni},
            pdfsubject={High Energy Physics - Phenomenology},
            pdfkeywords={Beyond the Standard Model, Dark Matter, Electroweak Phase Transition, Gravitational Waves, Higgs Portal, Multi-Component Dark Matter, Perturbative Unitarity, Relic Density, Vacuum Stability, Z4 Symmetry},
            pdfcreator={pdfLaTeX},
            pdfproducer={LaTeX with hyperref}}

\usetikzlibrary{matrix,shapes.geometric,decorations.pathmorphing,decorations.markings}

\settocbibname{References}

\arxivnumber{2606.11564}

\title{\boldmath Strong First-Order Electroweak Phase Transition and Gravitational Waves in a $\mathbb{Z}_4$ Fermion--Scalar Dark Matter Model}

\author{J. P. Carvalho-Corrêa\orcidlink{0009-0009-4420-771X},}
\author{J. P. Cunha-Melo\orcidlink{0009-0000-7561-5323},}
\author{I. M. Pereira\orcidlink{0000-0002-7203-6737},}
\author{B. L. Sánchez-Vega\orcidlink{0000-0002-6735-5813}}
\author{and A. C. D. Viglioni\orcidlink{0009-0005-1684-6419}}

\affiliation{Departamento de F\'isica, Universidade Federal de Minas Gerais (UFMG), \\
Av. Antônio Carlos 6627, Belo Horizonte, Minas Gerais, Brazil}

\emailAdd{jpcarv-15897@ufmg.br}
\emailAdd{jpcunha7l@ufmg.br}
\emailAdd{isaacmp@ufmg.br}
\emailAdd{bruce@ufmg.br}
\emailAdd{arthurcesar@ufmg.br}

\abstract{We investigate whether a minimal $\mathbb{Z}_4$-symmetric fermion--scalar extension of the Standard Model can simultaneously realise viable dark matter, a strong electroweak phase transition, and a stochastic gravitational-wave signal. The model contains a real scalar singlet and a Dirac fermion, allowing thermal two-component dark matter, mixed WIMP--FIMP histories, and an effectively fermionic relic abundance generated by scalar decays. We impose theoretical consistency, the correct electroweak vacuum, and dark-matter constraints from relic density, direct detection, and invisible Higgs decays before using the surviving points as input for the finite-temperature analysis. This reveals that the compatibility between dark matter and a strong first-order electroweak phase transition is highly selective. After current dark-matter constraints are imposed, the strong-transition criterion along the Higgs direction is satisfied only in two viable regimes: the thermal two-component case with $M_\psi<M_S<2M_\psi$ and the decay-driven WIMP--FIMP case with $M_S>2M_\psi$. By contrast, the thermal regime with $M_S<M_\psi$ and the stable mixed WIMP--FIMP scenario with $M_S<2M_\psi$ are largely concentrated at small portal couplings or near the Higgs-resonance region, and do not yield a strong transition in the parameter space considered. The successful transitions typically proceed through an intermediate singlet-like phase. For representative nucleating benchmark points in the viable strong-transition regions, we compute the gravitational-wave spectra from sound waves and turbulence. Some spectra overlap the projected sensitivity curves of future space-based interferometers, indicating potentially observable signals in selected dark-matter-compatible regions where a sufficiently active Higgs portal appears in correlated combination with the scalar mass and the remaining dark sector parameters.}

\begin{document}

\maketitle

\flushbottom

\section{Introduction}
\label{sec:introduction}

The Standard Model (SM) provides an accurate description of particle
interactions up to the electroweak scale, but it leaves open two central
questions in particle physics and cosmology. It contains no viable dark-matter
(DM) candidate, despite the robust evidence for a non-baryonic matter component
in the Universe \cite{Planck:2018vyg}. In addition, for the observed Higgs-boson
mass, the electroweak symmetry-breaking transition is a crossover rather than a
strong first-order transition, and therefore cannot support the usual
electroweak-baryogenesis mechanism \cite{Morrissey:2012db,Quiros:1999jp}. These facts motivate simple Higgs-portal extensions in which the fields
responsible for the dark sector also reshape the finite-temperature scalar
potential, allowing the origin of DM and the thermal history of electroweak
symmetry breaking to be studied within a common framework \cite{Mazumdar2019}.

A first-order cosmological phase transition would also provide a possible
source of a stochastic gravitational-wave (GW) background, potentially
accessible to future space-based interferometers such as LISA, DECIGO, and BBO
\cite{AthronWu,Caprini:2019egz,LISA:2017pwj,Kawamura:2021decigo,Crowder:2005bbo,
Schmitz:2020sjk}. This possibility makes dark sectors with nontrivial
electroweak thermal dynamics especially compelling. At the same time, the
required interactions are constrained by relic-density measurements,
direct-detection searches, invisible Higgs decays, and the existence of a
consistent electroweak vacuum at zero temperature. A realistic assessment must
therefore confront phase-transition dynamics with the same constraints that
shape the DM parameter space.

Singlet extensions of the SM provide a minimal setting in which this problem can
be addressed. Higgs-portal scalar sectors have been widely studied as
economical frameworks linking DM phenomenology, electroweak phase transitions,
and GW production
\cite{GonderingerLimRamseyMusolf,Cline:2013gha,Athron:2017Singlet,
Chiang:2018gsn,Vaskonen:2016yuv,Oikonomou:2024ewpt,Ghosh2025}. However, the cosmological
structure can change qualitatively once one goes beyond the minimal
$\mathbb{Z}_2$ stabilisation mechanism \cite{HuangMedinaWagner,BorahGhoshSaha,GuoKangZhang}. A $\mathbb{Z}_4$ symmetry allows
semi-annihilation, conversion processes among dark-sector species, and
decay-driven non-thermal production channels
\cite{DEramo:2010keq,Hall:2009bx,Feng:2003SuperWIMP_PRD,
Feng:2003SuperWIMP_PRL}. The resulting theories are not merely small
deformations of the scalar-singlet portal, but multi-channel dark sectors with
several thermal and non-thermal realisations
\cite{YagunaZapata2022,YagunaZapata2024}. More broadly, they fit into the
systematic study of multi-component dark sectors stabilised by discrete
symmetries \cite{CarvalhoCorrea:2025epjc}.

The dark-matter phenomenology of $\mathbb{Z}_4$ multi-component sectors has been
explored in recent work~\cite{YagunaZapata2022,YagunaZapata2024}. Here we
revisit this class of models under updated experimental constraints, especially
the stronger limits from current direct-detection searches~\cite{LZ:2024vif},
and investigate whether the dark-matter viable regimes can still
accommodate a strong electroweak phase
transition. This constitutes more than a straightforward extension of the
relic-density analysis. The Higgs portal controls both the scalar thermal
contact with the SM bath and the communication between the Higgs and singlet
directions in the finite-temperature potential. As a result, the same coupling
that can help generate a barrier between phases is also constrained by direct
detection whenever the scalar survives as a present-day dark-matter component.

In this work we focus on the minimal fermion--scalar realisation of this
mechanism: a real singlet and a Dirac fermion charged under a
stabilising $\mathbb{Z}_4$ symmetry. The scalar provides the Higgs-portal
connection to the SM plasma, while the $\mathbb{Z}_4$ Yukawa interactions
determine conversion, semi-annihilation, and non-thermal fermion production.
This separation of roles makes the model a useful framework for tracing how
dark-sector production mechanisms, direct searches, and electroweak thermal
dynamics constrain one another.

This setup leads to two questions. First, after imposing zero-temperature
consistency conditions and current dark-matter constraints, which cosmological
regimes retain sufficiently active scalar-sector dynamics to satisfy the
strong-transition criterion? Second, among those regimes, which transitions
nucleate successfully, and can the resulting gravitational-wave spectra fall
within the projected reach of future interferometers? These questions require
the dark-matter and finite-temperature analyses to be treated as connected
parts of the same problem, rather than as independent scans.

To address these questions, we follow a sequential filtering strategy. We first
determine the parameter regions compatible with boundedness from below,
perturbative unitarity, the correct electroweak vacuum, relic-density
constraints, direct-detection limits, and invisible Higgs decays. We then use
only these phenomenologically viable configurations as input for the one-loop
finite-temperature effective potential and the electroweak phase-transition
analysis. For representative benchmarks satisfying the strong-transition
criterion and allowing bubble nucleation, we compute the corresponding
stochastic gravitational-wave spectra from sound waves and turbulence in the
plasma. This procedure makes explicit how present dark-sector constraints
select the regions relevant for electroweak thermal dynamics and
gravitational-wave production.

The paper is organised as follows. In Sec.~\ref{sec:model}, we introduce the
$\mathbb{Z}_4$-symmetric fermion--scalar extension of the SM. In
Sec.~\ref{sec:constraints}, we discuss the zero-temperature consistency
conditions and the vacuum structure. Section~\ref{sec:dark_matter} is devoted
to the DM phenomenology and to the allowed cosmological regimes of the model.
In Sec.~\ref{sec:effective_potential}, we construct the one-loop effective
potential and its finite-temperature extension. The electroweak phase
transition is analysed in Sec.~\ref{sec:EWPT}, while the corresponding GW
spectra for representative benchmark configurations are discussed in
Sec.~\ref{sec:GW}. Finally, our conclusions are presented in
Sec.~\ref{sec:conclusions}.

\section{A \texorpdfstring{$\mathbb{Z}_4$}{Z4}-Symmetric Fermion--Scalar Extension of the SM}
\label{sec:model}

We consider a minimal fermion--scalar extension of the SM containing a real
scalar singlet $S$ and a Dirac fermion $\psi$, both neutral under the SM gauge
group. No new gauge interactions or SM-charged fields are introduced. The dark
sector is stabilised by a discrete $\mathbb{Z}_4$ symmetry, under which
\begin{equation}
\psi \;\to\; i\,\psi, \qquad
S \;\to\; -S, \qquad
\text{SM fields} \;\to\; \text{SM fields}.
\label{eq:z4_symmetry}
\end{equation}
Equivalently, the corresponding $\mathbb{Z}_4$ charges are $q_\psi = 1$ and
$q_S = 2 \pmod 4$. This assignment forbids odd powers of $S$, excludes the
standard Yukawa interaction $\bar{\psi}\psi\,S$, and allows the 
operator $(H^\dagger H)S^2$, where $H$ denotes the SM Higgs doublet. It also permits fermion bilinears involving
$\psi^c$, which generate conversion, semi-annihilation, and, when
kinematically open, scalar decays into fermion pairs.

The renormalisable Lagrangian consistent with this field content and symmetry
assignment is
\begin{equation}
\mathcal{L}
=
\mathcal{L}_{\rm SM}
+\frac{1}{2}\partial_\mu S\,\partial^\mu S
+i\bar{\psi}\slashed{\partial}\psi
- M_\psi \bar{\psi}\psi
-\frac{1}{2}\left[
y_s\,\overline{\psi^c}\psi
+
y_p\,\overline{\psi^c}\gamma_5\psi
+
\text{h.c.}
\right]S
- V_0(H,S).
\label{eq:lagrangian}
\end{equation}
 In the following, we restrict attention
to a CP-conserving setup and take the Yukawa couplings $y_s$ and $y_p$ to be
real.

The tree-level scalar potential is
\begin{equation}
V_0(H,S)
=
-\mu_H^2\,H^\dagger H
+\lambda_H (H^\dagger H)^2
-\frac{1}{2}\mu_S^2 S^2
+\frac{1}{4}\lambda_S S^4
+\frac{1}{2}\lambda_{HS}(H^\dagger H)S^2.
\label{eq:tree_potential}
\end{equation}
The Higgs portal $\lambda_{HS}$ is the only renormalisable interaction that
couples the singlet directly to the SM. Together with $\mu_S^2$ and
$\lambda_S$, it controls the zero-temperature scalar vacuum structure and affects the
finite-temperature effective potential.

At zero temperature, we focus on the parameter region in which the physical
vacuum preserves the dark symmetry,
\begin{equation}
\langle H\rangle
=
\frac{1}{\sqrt{2}}
\begin{pmatrix}
0\\ v
\end{pmatrix},
\qquad
\langle S\rangle = 0.
\end{equation}
The electroweak vacuum is therefore located at $(v,0)$ and the $\mathbb{Z}_4$
symmetry remains unbroken today. This ensures the stability of the dark sector,
although the unbroken symmetry still allows the decay $S\to\psi\psi$ when
$M_S>2M_\psi$. For $M_S<2M_\psi$, both dark-sector particles are stable. During
the thermal evolution, however, the potential may develop intermediate minima
with a non-vanishing singlet background, which can affect the phase-transition
dynamics.

At tree level, stationarity along the Higgs direction implies
\begin{equation}
\mu_H^2=\lambda_H v^2,
\qquad
M_h^2=2\lambda_H v^2.
\label{eq:higgs_tree_relations}
\end{equation}
The singlet mass at the electroweak vacuum is defined by the curvature of the
potential along the $S$ direction,
\begin{equation}
M_S^2
\equiv
\left.\frac{\partial^2 V_0}{\partial S^2}\right|_{(v,0)}
=
-\mu_S^2+\frac{1}{2}\lambda_{HS}v^2.
\label{eq:ms_tree}
\end{equation}
In the numerical analysis, we trade the Lagrangian parameter $\mu_S^2$ for the
zero-temperature input mass $M_S$.

In the physical vacuum, the dark fermion remains a Dirac state with mass
$M_\psi$. Away from this vacuum, a non-vanishing singlet background induces
Majorana-like entries in the fermionic sector through the Yukawa interactions
in Eq.~\eqref{eq:lagrangian}. The fermionic spectrum is therefore
background-dependent, which is relevant for the one-loop effective potential
and for the thermal history of the model.

After expressing the Higgs-sector parameters in terms of $v$ and $M_h$, the new
sector is described by
\begin{equation}
\{ M_S,\; M_\psi,\; \lambda_S,\; \lambda_{HS},\; y_s,\; y_p \}.
\end{equation}
The scalar couplings determine the vacuum structure and the finite-temperature
potential, while the portal and Yukawa interactions control the dark-matter
production channels. Depending on the hierarchy between $M_S$ and $M_\psi$, the
model realises either a two-component present-day dark sector, when
$M_S<2M_\psi$, or an effectively one-component fermionic relic abundance when
the scalar decay $S\to\psi\psi$ is open
\cite{YagunaZapata2022,YagunaZapata2024}. The theoretical constraints on this
parameter space are discussed next.

\section{Zero-Temperature Consistency Conditions}
\label{sec:constraints}

We now define the theoretically consistent zero-temperature region of parameter
space used in the phenomenological and thermal analyses. Three requirements are
imposed at tree level: boundedness from below of the scalar potential,
perturbative unitarity of scalar scattering amplitudes, and the condition that
the electroweak vacuum identified in Sec.~\ref{sec:model} be the physical
vacuum today
\cite{Kannike:2012rw,Lee:1977eg,Lee:1977yc,Logan:2022mru}. These conditions
fix the scalar-sector domain over which the dark-matter and phase-transition
analyses are performed.

\subsection{Boundedness from Below}

The potential must be bounded from below in the large-field regime. This
requirement is controlled by the quartic part of the tree-level potential,
\begin{equation}
V_0^{(4)}(h,s)=\frac{\lambda_H}{4}h^4+\frac{\lambda_S}{4}s^4+\frac{\lambda_{HS}}{4}h^2s^2.
\end{equation}
For a two-field quartic form, boundedness from below is equivalent to the
copositivity conditions \cite{Kannike:2012rw}
\begin{equation}
\lambda_H>0, \qquad
\lambda_S>0, \qquad
\lambda_{HS}+2\sqrt{\lambda_H\lambda_S}>0.
\label{eq:bfb_conditions}
\end{equation}
These inequalities exclude runaway directions in the scalar field space.

\subsection{Perturbative Unitarity}

We also require tree-level perturbative unitarity in scalar $2\to2$ scattering.
The quartic couplings are constrained by the eigenvalues of the coupled scalar
scattering matrix. Imposing the standard zeroth-partial-wave condition
\begin{equation}
|a_0|\leq \frac{1}{2},
\end{equation}
one obtains \cite{Lee:1977eg,Lee:1977yc,Goodsell:2018tti}
\begin{equation}
|\lambda_H|\leq 4\pi, \qquad
|\lambda_{HS}|\leq 8\pi, \qquad
|\Lambda_\pm|\leq 8\pi,
\label{eq:unitarity_bounds}
\end{equation}
where
\begin{equation}
\Lambda_\pm
\equiv
\frac{1}{2}\left(
6\lambda_H+3\lambda_S
\pm
\sqrt{36\lambda_H^2+9\lambda_S^2+4\lambda_{HS}^2-36\lambda_H\lambda_S}
\right).
\label{eq:Lambda_pm}
\end{equation}

These bounds provide upper limits on the scalar quartic couplings and, in
particular, restrict how far the singlet self-coupling can be increased above
the lower value required by the vacuum structure. Further details are given in
Appendix~\ref{app:unitarity_details}.

\subsection{Electroweak Vacuum and Viable Scalar Parameter Space}
\label{subsec:vacuum_structure}

In addition to boundedness from below and perturbative unitarity, we require the
electroweak vacuum $(v,0)$ to be the physical zero-temperature vacuum. This
condition ensures that electroweak symmetry is broken while the
$\mathbb{Z}_4$ symmetry remains intact in the present Universe.

Using the tree-level potential in the neutral background-field directions,
\begin{equation}
    V_0(h,s)=-\frac{1}{2}\mu_H^2 h^2+\frac{1}{4}\lambda_H h^4-\frac{1}{2}\mu_S^2 s^2+\frac{1}{4}\lambda_S s^4+\frac{1}{4}\lambda_{HS}h^2 s^2,
\end{equation}
together with the tree-level relations ~\eqref{eq:higgs_tree_relations} and \eqref{eq:ms_tree} introduced in Sec.~\ref{sec:model}, $\mu_H^2 = \lambda_H v^2$ and $M_S^2 = -\mu_S^2 + \frac{1}{2}\lambda_{HS}v^2$, we express the vacuum conditions in terms of the physical input $M_S$ and the portal coupling $\lambda_{HS}$.

The potential may contain stationary points away from $(v,0)$, including a
pure singlet configuration $(0,v_s)$ or a mixed configuration with both
backgrounds nonzero. As shown in Appendix~\ref{app:vacuum_structure_details},
once the electroweak vacuum is required to be locally stable, namely
\begin{equation}
M_S^2>0,
\end{equation}
the mixed stationary point does not provide an independent competing minimum:
it is either absent as a physical solution or corresponds to a saddle point.
The only relevant tree-level competitor to the electroweak vacuum is therefore
the pure singlet extremum.

If $\mu_S^2>0$, the potential admits a singlet-breaking stationary point,
\begin{equation}
(0,v_s),
\qquad
v_s^2=\frac{\mu_S^2}{\lambda_S},
\end{equation}
with vacuum energy \(V_0(0,v_s)=-\mu_S^4/(4\lambda_S)\). Since
\(V_0(v,0)=-\mu_H^4/(4\lambda_H)\), requiring the electroweak vacuum to be
deeper gives
\begin{equation}
V_0(v,0)<V_0(0,v_s)
\quad\Longrightarrow\quad
\lambda_S>\lambda_H\,\frac{\mu_S^4}{\mu_H^4}.
\label{eq:vacuum_depth_condition_main}
\end{equation}
Using Eqs.~\eqref{eq:higgs_tree_relations} and \eqref{eq:ms_tree}, this condition becomes
\begin{equation}
\lambda_S>\lambda_S^{\rm min},
\qquad
\lambda_S^{\rm min}
=
\frac{\left(M_S^2-\frac{1}{2}\lambda_{HS}v^2\right)^2}{\lambda_H v^4}.
\label{eq:lambdaSmin_main}
\end{equation}
For $\mu_S^2\leq 0$, the pure singlet extremum is absent and this additional
vacuum-depth bound does not apply.

It is useful to encode this vacuum requirement in a point-dependent reference
value for the singlet quartic coupling,
\begin{equation}
\lambda_S^{\rm base}
=
\begin{cases}
\lambda_S^{\rm min}, & \mu_S^2>0,\\[1mm]
0, & \mu_S^2\leq 0.
\end{cases}
\label{eq:lambdaSbase_main}
\end{equation}

The meaning of $\lambda_S^{\rm base}$ is simple: when a singlet-breaking
extremum exists, it is the lower value required for the electroweak vacuum to
be deeper than that extremum; when no such extremum exists, no additional
vacuum-depth lower bound is present. We then write
\begin{equation}
\lambda_S=\lambda_S^{\rm base}+a.
\label{eq:lambdaS_scan_main}
\end{equation}

The offset $a$ parametrises how far the singlet self-coupling is chosen above
the vacuum-preserving lower value. It is not an unconstrained parameter. For
each point in the $(M_S,\lambda_{HS})$ plane, we determine the maximum allowed
value $a_{\rm max}(M_S,\lambda_{HS})$ from the
perturbative-unitarity constraints. We then sample $a$ randomly in the interval

\begin{equation}
0<a<a_{\rm max}(M_S,\lambda_{HS}).
\label{eq:a_scan_interval}
\end{equation}

Each sampled point is subsequently required to satisfy the
boundedness-from-below conditions in Eq.~\eqref{eq:bfb_conditions} and is
rejected otherwise. Equivalently, the retained values of
$\lambda_S$ {\color{blue}lie} only within the theoretically allowed interval
above $\lambda_S^{\rm base}$. This prescription separates the lower bound
required by the zero-temperature vacuum structure from the upper bounds
imposed by perturbative unitarity.

\begin{figure}[!ht]
    \centering
    \includegraphics[width=0.88\textwidth]{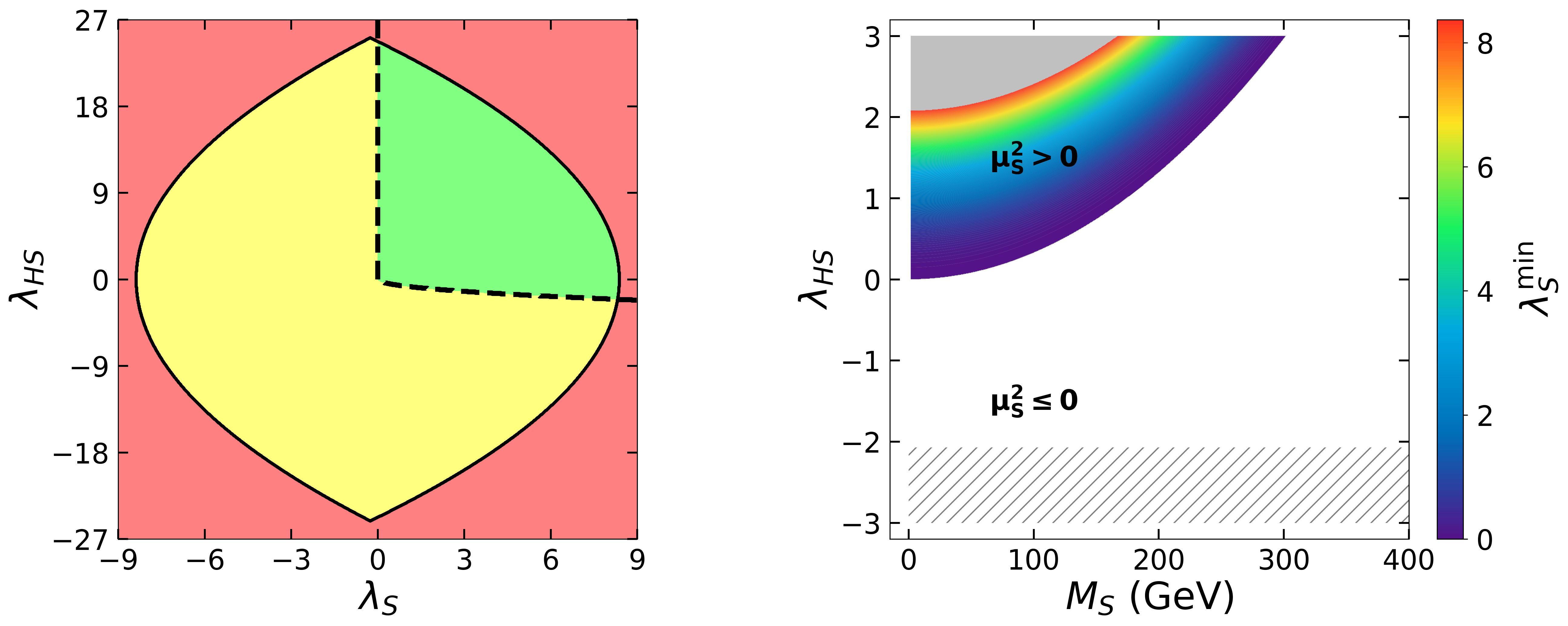}
   \caption{
Theoretical constraints on the scalar sector. 
Left: viable parameter space in the $(\lambda_S,\lambda_{HS})$ plane. 
The dashed curves mark the bounded-from-below boundaries. 
The green region satisfies both boundedness-from-below and perturbative-unitarity requirements, whereas the yellow region remains perturbatively unitary but is excluded by boundedness from below. 
Right: lower bound $\lambda_S^{\rm min}$ imposed by the possible singlet-breaking extremum as a function of the singlet mass $M_S$, for representative values of the Higgs-portal coupling $\lambda_{HS}$. 
Colour-filled regions correspond to $\mu_S^2>0$, while white regions correspond to $\mu_S^2\leq 0$. 
The grey region is excluded by the combined requirements of perturbative unitarity and an electroweak vacuum that is the global minimum. 
The lower hatched region is ruled out by boundedness-from-below and unitarity constraints on the Higgs-portal coupling.
}
    \label{fig:Z422}
\end{figure}

The combined impact of these requirements is illustrated in
Fig.~\ref{fig:Z422}. The left panel shows the allowed region in the
$(\lambda_S,\lambda_{HS})$ plane after imposing boundedness from below and
perturbative unitarity. The right panel shows the lower value
$\lambda_S^{\rm min}$ required by the possible singlet-breaking extremum as a
function of \(M_S\), for representative values of the Higgs-portal coupling.
Together, these constraints define the scalar-sector domain used in the
dark-matter scan and in the finite-temperature analysis.

\section{Dark Matter Phenomenology and Viable Cosmological Regimes}
\label{sec:dark_matter}

\begin{figure}[htbp]
    \centering
    \includegraphics[width=1.0\textwidth]{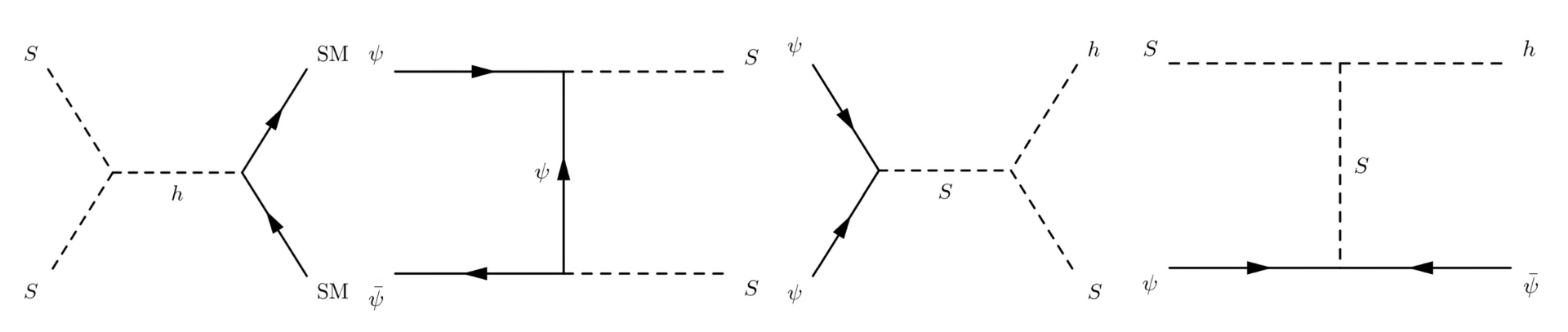}
    \caption{
Representative dark-sector annihilation, conversion, semi-annihilation, and
co-scattering processes. From left to right:
$SS\to h^\ast\to \mathrm{SM}\,\overline{\mathrm{SM}}$,
$\psi\bar\psi\to SS$,
$\psi\psi\to Sh$, and
$S\psi\to h\bar\psi$.
The arrows indicate fermion-number flow. Vertices with two incoming or two
outgoing arrows arise from the interaction
$S\overline{\psi^c}(y_s+y_p\gamma_5)\psi+\mathrm{h.c.}$
Charge-conjugate processes are implicit. For $\psi\bar\psi\to SS$, the crossed
contribution obtained by exchanging the two identical final-state scalars is
also understood.
}
    \label{fig:diagramZ_all}
\end{figure}

The dark sector contains a Dirac fermion $\psi$ and a real scalar $S$, both
stabilised by the $\mathbb{Z}_4$ symmetry. Its phenomenology is governed by the
Higgs portal $\lambda_{HS}$, which keeps the scalar in contact with the SM bath,
and by the Yukawa couplings $y_s$ and $y_p$, which mediate dark-sector
conversion, semi-annihilation, and, when kinematically allowed, scalar decay.
Representative processes are shown in Fig.~\ref{fig:diagramZ_all}. These
$\mathbb{Z}_4$-specific channels distinguish the model from the usual
$\mathbb{Z}_2$ scalar portal~\cite{YagunaZapata2022,DEramo:2010keq}.

Although $\psi$ is a Dirac field at the physical vacuum, the Yukawa operators
containing $\psi^c$ change the continuous Dirac fermion number by two units,
while preserving the $\mathbb{Z}_4$ symmetry. Amplitudes containing these
vertices are evaluated using the continuous auxiliary fermion-flow
prescription of Refs.~\cite{DennerPLB,DennerNPB}. In the FeynRules
implementation, $\psi$ is defined as a non-self-conjugate Dirac field, while
the charge-conjugated and Hermitian-conjugate interactions are included
explicitly, following the treatment of conjugated fermion fields described in
Ref.~\cite{FeynRules}. Defining
\begin{equation}
  \Gamma=y_s+y_p\gamma_5,
  \qquad
  \bar\Gamma=\gamma^0\Gamma^\dagger\gamma^0,
\end{equation}
the corresponding reversed vertex matrices satisfy
\begin{equation}
  C\Gamma^T C^{-1}=\Gamma,
  \qquad
  C\bar\Gamma^T C^{-1}=\bar\Gamma.
\end{equation}
The two possible orientations of the auxiliary fermion flow therefore give
equivalent representations of the same fermion chain and are not separate
contributions to be summed.

This interaction pattern leads to two distinct roles for the dark fields. The
scalar undergoes Higgs-portal-driven freeze-out, while the fermion can behave
either as a WIMP or as a FIMP, depending on the size of the Yukawa couplings.
The portal interaction is therefore central not only to the scalar relic
density, but also to direct detection and to the finite-temperature scalar
potential.

The late-time composition is fixed by the mass hierarchy. For
$M_S<2M_\psi$, the decay $S\to\psi\psi$ is forbidden and both species are
stable, so that the present relic abundance is shared between $S$ and $\psi$.
For $M_S>2M_\psi$, the scalar decays into fermions and the present-day dark
matter is purely fermionic, although the final abundance can still receive a
SuperWIMP contribution from the thermal scalar relic
~\cite{Feng:2003SuperWIMP_PRD,Feng:2003SuperWIMP_PRL}.

We confront these possible histories with relic-density, direct-detection, and
invisible Higgs-decay constraints. The surviving points are then used as input
for the finite-temperature phase-transition analysis.

\subsection{Production Mechanisms and Relic Abundance}
\label{subsec:dm_production}

We now describe the Boltzmann system used to compute the relic abundances of
$S$ and $\psi$. The relevant cosmological histories are controlled by the
$S$--$\psi$ mass hierarchy and by the size of the Yukawa couplings. In the
regions considered below, the scalar remains thermally connected to the SM bath
through the Higgs portal, while the fermion may either thermalise with the dark
sector or be produced out of equilibrium through freeze-in~\cite{Hall:2009bx}.

\subsubsection{Coupled Boltzmann Equations}
\label{subsec:dark_matterBoltzmann}

We track the cosmological evolution of the comoving abundances
\begin{equation}
Y_\psi \equiv \frac{n_\psi+n_{\bar\psi}}{s},
\qquad
Y_S \equiv \frac{n_S}{s},
\qquad
x \equiv \frac{M_S}{T},
\end{equation}
where $s$ is the entropy density. We assume vanishing particle--antiparticle
asymmetries in the dark sector, so that $Y_\psi$ fully characterises the
fermionic abundance. Since the scalar is the state directly connected to the SM
thermal bath, we use $M_S$ as the reference scale in the variable $x$.

For compactness, we express the Boltzmann equations in terms of equilibrium
reaction densities, which include the relevant symmetry factors and the sum over
CP-conjugate channels. Defining
\begin{equation}
\bar Y_i \equiv Y_i^{\rm eq},
\end{equation}
the reaction densities are
\begin{equation}
\begin{aligned}
\bar\gamma_{\rm ann} &\equiv s^2 \bar Y_S^{\,2}\, \langle \sigma v \rangle_{SS\to {\rm SM}}, &
\bar\gamma_{\rm conv} &\equiv s^2 \bar Y_\psi^{\,2}\, \langle \sigma v \rangle^{\rm eff}_{\psi\bar\psi \leftrightarrow SS}, &
\bar\gamma_{\rm SA} &\equiv s^2 \bar Y_\psi^{\,2}\, \langle \sigma v \rangle^{\rm eff}_{\psi\psi \leftrightarrow Sh}, \\[1mm]
\bar\gamma_{\rm cos} &\equiv s^2 \bar Y_S \bar Y_\psi\, \langle \sigma v \rangle^{\rm eff}_{S\psi \leftrightarrow \bar\psi h}, &
\bar\gamma_{\rm dec} &\equiv s\,\bar Y_S\, \langle \Gamma_{S\to\psi\psi} \rangle, &
\langle \Gamma_{S\to\psi\psi} \rangle &= \Gamma_S^{\rm tot}\,\frac{K_1(x)}{K_2(x)}.
\end{aligned}
\end{equation}
Here, $K_n(x)$ denotes the modified Bessel function of the second kind. The
densities $\bar\gamma_{\rm ann}$, $\bar\gamma_{\rm conv}$,
$\bar\gamma_{\rm SA}$, $\bar\gamma_{\rm cos}$, and $\bar\gamma_{\rm dec}$
correspond to annihilation, conversion, semi-annihilation, co-scattering, and
decay, respectively. The semi-annihilation density includes both
$\psi\psi\to Sh$ and the CP-conjugate process $\bar\psi\bar\psi\to Sh$.

We also introduce the departure-from-equilibrium combinations
\begin{equation}
\Delta_{\rm ann} \equiv \frac{Y_S^2}{\bar Y_S^{\,2}} - 1,\
\Delta_{\rm conv} \equiv \frac{Y_\psi^2}{\bar Y_\psi^{\,2}} - \frac{Y_S^2}{\bar Y_S^{\,2}},\
\Delta_{\rm SA} \equiv \frac{Y_\psi^2}{\bar Y_\psi^{\,2}} - \frac{Y_S}{\bar Y_S},\ 
\Delta_{\rm dec} \equiv \frac{Y_S}{\bar Y_S} - \frac{Y_\psi^2}{\bar Y_\psi^{\,2}}, \
\Delta_{\rm cos} \equiv \frac{Y_S}{\bar Y_S} - 1.
\end{equation}
The coupled Boltzmann equations then take the form
\begin{align}
\frac{dY_\psi}{dx}
&=
-\frac{2}{sHx}
\left(
\bar\gamma_{\rm conv}\,\Delta_{\rm conv}
+
\bar\gamma_{\rm SA}\,\Delta_{\rm SA}
-
\bar\gamma_{\rm dec}\,\Delta_{\rm dec}
\right),
\label{eq:boltz_psi_final}
\\[2mm]
\frac{dY_S}{dx}
&=
-\frac{1}{sHx}
\left(
2\,\bar\gamma_{\rm ann}\,\Delta_{\rm ann}
-
2\,\bar\gamma_{\rm conv}\,\Delta_{\rm conv}
-
\bar\gamma_{\rm SA}\,\Delta_{\rm SA}
+
\bar\gamma_{\rm dec}\,\Delta_{\rm dec}
+
\bar\gamma_{\rm cos}\,\Delta_{\rm cos}
\right),
\label{eq:boltz_S_final}
\end{align}
where $H\equiv H(T)$ is the Hubble expansion rate.

The terms in Eqs.~\eqref{eq:boltz_psi_final} and
\eqref{eq:boltz_S_final} have a direct particle-number interpretation.
Annihilation depletes the scalar abundance, conversion transfers number density
between the scalar and fermionic sectors, semi-annihilation changes both
abundances, and scalar decay removes one scalar while producing two fermionic
quanta. In the CP-symmetric limit adopted here, the co-scattering process
$S\psi\leftrightarrow\bar\psi h$ affects the scalar equation but does not
change the total fermionic abundance encoded in $Y_\psi$.

In the FIMP regime,
\begin{equation}
Y_\psi \ll \bar Y_\psi ,
\end{equation}
the fermion never reaches thermal equilibrium. Its final abundance is generated
out of equilibrium through inverse conversion, inverse semi-annihilation, and,
when the channel is open, through the decay $S\to\psi\psi$, rather than by
conventional freeze-out.

\subsubsection{Viable Cosmological Regimes}
\label{subsec:dark_matterProduction}

The four regimes considered in this work are defined by two inputs: the
$S$--$\psi$ mass hierarchy and the size of the Yukawa couplings. The mass
hierarchy determines whether the scalar is stable or unstable, while the
Yukawa couplings determine whether the fermion is thermally produced or remains
out of equilibrium. The resulting classification is summarised in
Table~\ref{tab:scenarios}.

Scenarios~I and II correspond to thermal two-component dark matter. In both
cases $S$ and $\psi$ are stable, but the ordering of $M_S$ and $M_\psi$ affects
the relic composition and the impact of direct detection, since only the scalar
couples to nucleons at tree level through Higgs exchange. Scenarios~III and IV
instead involve a feebly coupled fermion. In Scenario~III, $M_S<2M_\psi$ and
both particles remain stable today. In Scenario~IV, $M_S>2M_\psi$, so the scalar
decays into fermions and the late-time abundance is purely fermionic.

\begin{table}[htbp]
\centering
\renewcommand{\arraystretch}{1.4}
\caption{Cosmological regimes considered in the scan. The WIMP/FIMP labels
refer to the dominant production mechanism of $(\psi,S)$, respectively, while
the late-time content indicates which species remain as present-day dark
matter.}
\begin{tabular}{cccc}
\hline
\textbf{Scenario} & \textbf{Production mode $(\psi,S)$} & \textbf{Late-time content} & \textbf{Mass hierarchy} \\
\hline
\multicolumn{4}{l}{\itshape Thermal two-component regimes} \\
I & WIMP + WIMP & $\psi + S$ stable & $M_\psi<M_S<2M_\psi$ \\
II & WIMP + WIMP & $\psi + S$ stable & $M_S<M_\psi$ \\
\hline
\multicolumn{4}{l}{\itshape Mixed thermal / non-thermal regimes} \\
III & FIMP + WIMP & $\psi + S$ stable & $M_S<2M_\psi$ \\
IV & FIMP + WIMP & only $\psi$ survives & $M_S>2M_\psi$ \\
\hline
\end{tabular}
\label{tab:scenarios}
\end{table}

\subsubsection{Relic-Density Criterion}
\label{subsec:relic_density_criterion}

The relic abundance is computed with \texttt{micrOMEGAs}~6.0~\cite{micromegas},
using the coupled Boltzmann system for multi-component dark matter, including
annihilation, conversion, semi-annihilation, co-scattering, and scalar decay
whenever the corresponding channels are kinematically open. A point is retained
if the total dark-matter abundance satisfies
\[
\bigl|\Omega_{\rm DM}h^2 - 0.12\bigr| \le 0.012.
\]
This $10\%$ window is used as a phenomenological acceptance criterion for the
scan, allowing for numerical and theoretical uncertainties in the computation
of the relic abundance, rather than as the experimental uncertainty on the
Planck measurement~\cite{Planck:2018vyg}. In two-component regimes,
\[
\Omega_{\rm DM}h^2=\Omega_\psi h^2+\Omega_S h^2,
\]
whereas in the scalar-decay regime the late-time abundance is purely
fermionic. In this regime, we additionally require the scalar lifetime to be
short enough not to disturb standard BBN. We impose the conservative condition
\[
\tau_S \lesssim 1~{\rm s},
\]
so that the scalar decays before the onset of standard light-element
nucleosynthesis. When both species are stable, we define the fractional abundances
\[
\xi_i\equiv\frac{\Omega_i}{\Omega_{\rm DM}},
\qquad i=\psi,S,
\]
which enter the rescaling of the direct-detection rates.

\subsection{Phenomenological Constraints}
\label{subsec:experimental_constraints}

We further constrain the relic-density-compatible points using
spin-independent direct detection and, for sufficiently light scalars, invisible
Higgs decays. These probes act differently across the four regimes. When both
dark-sector particles are stable, the direct-detection rate receives
contributions from each component weighted by its relic fraction. When the
scalar decays, only the surviving fermionic component contributes to
present-day direct-detection signals.

\subsubsection{Direct Detection}

\begin{figure}[htbp]
    \centering
    \includegraphics[width=0.6\textwidth]{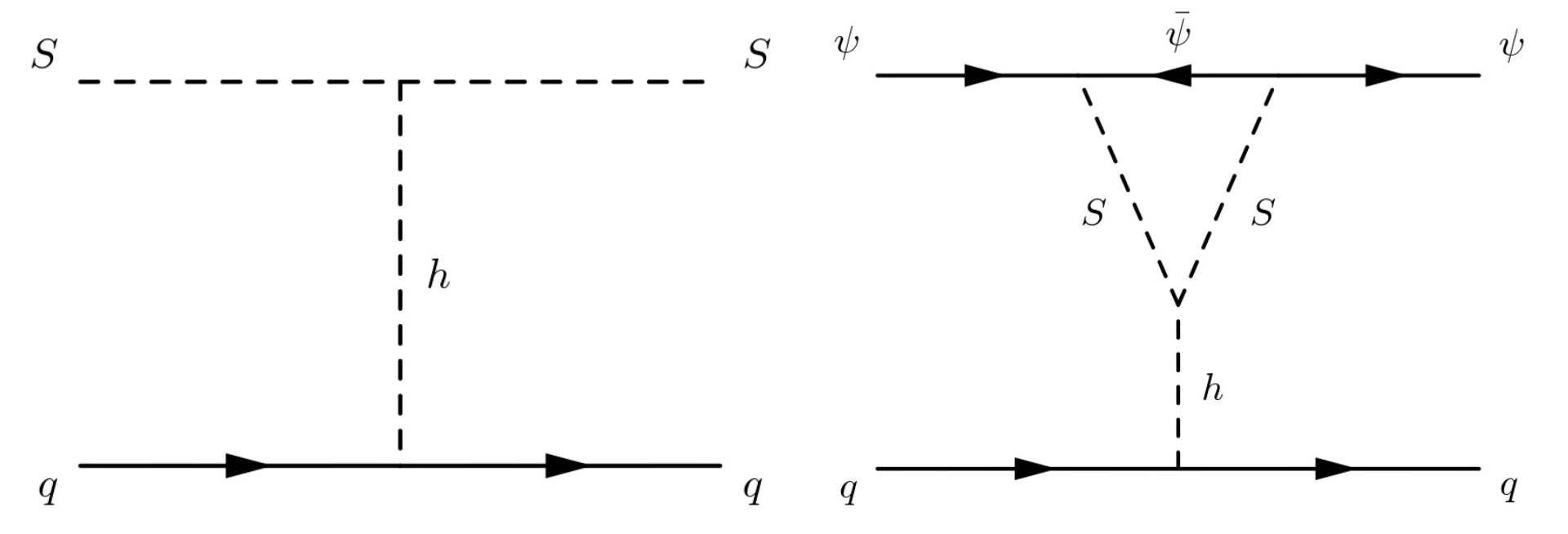}
    \caption{Diagrams for dark matter--nucleus scattering. Left: tree-level Higgs exchange for scalar dark matter. Right: loop-induced Higgs exchange for fermionic dark matter. Solid black lines denote nucleons, while dashed lines denote mediators.}
    \label{fig:detecdire}
\end{figure}

Direct-detection bounds constrain the model through spin-independent (SI)
scattering off nuclei, illustrated in Fig.~\ref{fig:detecdire}. The scalar
$S$ couples to nucleons at tree level through Higgs exchange induced by
$\lambda_{HS}$. The fermion $\psi$ has no tree-level coupling to the Higgs in
the physical vacuum, since $\langle S\rangle=0$, and scatters only through
loop-induced Higgs exchange involving $S$ and the Yukawa couplings $y_s$ and
$y_p$~\cite{YagunaZapata2022}.

For the scalar, the zero-momentum SI cross section is
\[
\sigma_{S}^{\rm SI} \simeq
\frac{\lambda_{HS}^2 f_N^2}{4\pi}
\frac{\mu_{SN}^{2} m_{N}^{2}}{M_{h}^{4} M_{S}^{2}},
\qquad
\mu_{SN} = \frac{M_S m_N}{M_S+m_N},
\]
where $f_N\simeq 0.30$ is the effective nucleon form factor.

For the fermion, the loop-induced SI cross section reads~\cite{YagunaZapata2022}
\[
\sigma_{\psi}^{\rm SI}
=
\frac{1}{\pi}
\frac{\mu_{\psi N}^2 m_N^2 f_N^2}{M_h^4}
\left[
\frac{\lambda_{HS}}{16\pi^2 M_\psi}
\bigl(|y_s|^2 f(r) + |y_p|^2 g(r)\bigr)
\right]^2,
\]
with
\[
r=\frac{M_S^2}{M_\psi^2},
\qquad
\mu_{\psi N}=\frac{M_\psi m_N}{M_\psi+m_N},
\]
and
\begin{align*}
f(r) &=
\frac{r^2-5r+4}{2\sqrt{r(4-r)}}
\arctan\!\Bigl(\frac{\sqrt{4-r}}{\sqrt{r}}\Bigr)
+\frac{1}{2}\bigl[2-(r-3)\ln r\bigr],
\\
g(r) &=
\frac{(r-3)\sqrt{r}}{2\sqrt{4-r}}
\arctan\!\Bigl(\frac{\sqrt{4-r}}{\sqrt{r}}\Bigr)
+\frac{1}{2}\bigl[2-(r-1)\ln r\bigr].
\end{align*}
For $r>4$, the loop functions are understood by analytic continuation.

In two-component regimes, we assume that the local density fractions follow the
cosmological fractions,
\[
\xi_i\equiv \frac{\Omega_i}{\Omega_{\rm DM}},
\qquad i=S,\psi.
\]
We impose the latest LZ spin-independent bound~\cite{LZ:2024vif} through the
practical recast
\begin{align}
\label{eq:dirdet}
\sum_{i=S,\psi}
\frac{\xi_i\,\sigma_i^{\rm SI}}{\sigma_{\rm limit}(M_i)}
<1,
\end{align}
where $\sigma_{\rm limit}(M_i)$ denotes the corresponding $90\%$ C.L.
single-component LZ upper limit at mass $M_i$. This criterion constrains the
total predicted recoil rate after relic-fraction rescaling. It should not be
interpreted as requiring each component to satisfy the single-component limit
independently.

When $M_S>2M_\psi$, the scalar is unstable and does not contribute to the
present-day halo abundance. In Scenario~III, both components are stable but the
fermion is FIMP-like; its loop-induced scattering rate is negligible, so direct
detection effectively constrains the scalar WIMP fraction. In Scenario~IV,
only the feebly coupled fermion survives today, and the direct-detection rate is
typically negligible. This difference is important for the finite-temperature
analysis, because direct-detection limits force the Higgs portal to be small
only when the scalar remains stable at late times.

\textit{Indirect detection.} We do not impose indirect-detection constraints in
the present analysis. The late-time signals in this model need not map directly
onto the standard two-body annihilation channels into SM final states usually
assumed in experimental limits. In particular, semi-annihilation processes such
as $\psi\psi\to Sh$ lead to final states whose interpretation depends on the
subsequent Higgs decay products, on whether the scalar is stable or decays into
fermions, and on the relative abundances of the stable dark-sector components.
A robust comparison with gamma-ray or cosmic-ray searches would therefore
require a dedicated spectral analysis. Since Ref.~\cite{YagunaZapata2022} found
indirect-detection bounds to have only a marginal impact on the viable
parameter space, we leave such an analysis for future work.

\subsubsection{Invisible Higgs Decay}

For $M_S<M_h/2$, the decay $h\to SS$ is kinematically allowed, with partial
width
\[
\Gamma_{h\to SS}
=
\frac{\lambda_{HS}^2 v^2}{32\pi M_h}
\sqrt{1-\frac{4M_S^2}{M_h^2}} .
\]
We require
\[
\mathrm{BR}(h\to\mathrm{inv})
=
\frac{\Gamma_{h\to SS}}
{\Gamma_h^{\rm SM}+\Gamma_{h\to SS}}
\leq 0.107,
\]
following the ATLAS Run 1+2 limit at $95\%$ C.L.~\cite{ATLAS:2023InvComb}.
This constraint applies whenever the singlet-pair channel is open and the
resulting final state is invisible at collider scales. In the stable-scalar
regimes, this corresponds directly to $h\to SS$ with missing energy. In the
scalar-decay regime, it corresponds to the invisible cascade
$h\to SS\to 4\psi$. The bound is therefore especially relevant for light
scalars with sizeable Higgs-portal couplings.

\subsection{Scan Setup and Viable Parameter Space}
\label{subsec:parameter_scan}

We now specify the parameter ranges used in the numerical exploration of the
$\mathbb{Z}_4$ model. For the dark-matter analysis, the scan is organised in
terms of
\begin{equation}
\{ M_S,\; M_\psi,\; \lambda_{HS},\; y_s,\; y_p \},
\end{equation}
while the SM inputs are fixed to their PDG values~\cite{PDGDM}.

The singlet quartic coupling $\lambda_S$ is not treated as an independent
dark-matter scan parameter. At tree level, the relic abundance, scalar decay,
and direct-detection rates are controlled by the masses, the Higgs portal, and
the Yukawa couplings; $\lambda_S$ affects instead the singlet self-interaction,
the zero-temperature vacuum structure, and the finite-temperature potential.
When a value of $\lambda_S$ is required, it is assigned point by point using
the vacuum-preserving prescription of Eq.~\eqref{eq:lambdaS_scan_main}. The
offset entering this prescription is sampled within the theoretically allowed
interval determined by boundedness from below, perturbative unitarity, and the
adopted perturbativity range for scalar quartics, as described in
Sec.~\ref{subsec:vacuum_structure}. Points for which this interval is empty are
discarded.

The scan covers the four cosmological regimes described in
Sec.~\ref{subsec:dark_matterProduction}. Masses and coupling magnitudes are
sampled logarithmically over the intervals listed below. For the dimensionless
couplings that may take either sign, namely $\lambda_{HS}$, $y_s$, and $y_p$,
we sample the absolute value logarithmically and assign the sign independently.
This allows both positive and negative portal interactions, as well as different
relative signs between the scalar and pseudoscalar Yukawa couplings. As far as
possible, the scan ranges follow Refs.~\cite{YagunaZapata2022,YagunaZapata2024},
facilitating comparison with previous studies of minimal $\mathbb{Z}_4$ dark
sectors.

For the Higgs-portal coupling, we consider
\begin{equation}
|\lambda_{HS}| \in [10^{-4},\,3].
\end{equation}
This interval includes small-portal regions relevant for direct-detection
survival and Higgs-resonance annihilation, as well as sizeable portal couplings
that can affect the finite-temperature scalar potential.

For the Yukawa couplings, we distinguish WIMP-like and FIMP-like fermion
production. In the WIMP-like case, we scan
\begin{equation}
|y_s|,\;|y_p| \in [10^{-2},\,3],
\end{equation}
so that the fermion can thermalise with the dark sector and participate in
freeze-out dynamics. In the FIMP-like case, we take
\begin{equation}
|y_s|,\;|y_p| \in [10^{-13},\,10^{-5}],
\end{equation}
so that $\psi$ remains out of equilibrium and is produced through freeze-in
processes mediated by the scalar sector.

The mass ranges are chosen according to the four regimes summarised in
Table~\ref{tab:scan_params}.

\begin{table}[htbp]
\centering
\footnotesize
\renewcommand{\arraystretch}{1.35}
\caption{Parameter ranges adopted in the numerical scan. Masses are given in
GeV. Coupling magnitudes are sampled logarithmically in the intervals shown,
with signs assigned independently for $\lambda_{HS}$, $y_s$, and $y_p$. The
Yukawa intervals distinguish WIMP-like and FIMP-like fermion production.}
\begin{tabularx}{\textwidth}{@{}>{\raggedright\arraybackslash}p{0.34\textwidth}>{\raggedright\arraybackslash}p{0.33\textwidth}>{\raggedright\arraybackslash}X@{}}
\hline
\textbf{Parameter / Regime} & \textbf{Scanning range} & \textbf{Role} \\
\hline
$|\lambda_{HS}|$ & $[10^{-4},\,3]$ & Higgs portal controlling scalar thermalisation, direct detection, and the finite-temperature scalar potential\\
$|y_s|,\;|y_p|$ (WIMP-like $\psi$) & $[10^{-2},\,3]$ & Thermal production, conversion, and semi-annihilation involving $\psi$ \\
$|y_s|,\;|y_p|$ (FIMP-like $\psi$) & $[10^{-13},\,10^{-5}]$ & Out-of-equilibrium fermion production \\
\hline
Scenario I: $M_\psi<M_S<2M_\psi$ & $M_\psi\in[40,\,2000],\; M_S\in[M_\psi,\,2M_\psi]$ & Two-component thermal regime \\
Scenario II: $M_S<M_\psi$ & $M_S\in[40,\,2000],\; M_\psi\in[M_S,\,3M_S]$ & Two-component thermal regime \\
Scenario III: $M_S<2M_\psi$ & $M_\psi\in[0.1,\,10000],\; M_S\in[0.1,\,2M_\psi]$ & Mixed two-component WIMP--FIMP regime \\
Scenario IV: $M_S>2M_\psi$ & $M_\psi\in[0.1,\,2000],\; M_S\in[2M_\psi,\, 8M_\psi]$ & Effectively one-component WIMP--FIMP regime with scalar decay \\
\hline
\end{tabularx}
\label{tab:scan_params}
\end{table}

\subsection{Scenario-by-Scenario Analysis}
\label{subsec:dm_results}

We now examine the parameter regions that survive the full set of theoretical
and phenomenological constraints. Each scanned point is first required to
satisfy the zero-temperature consistency conditions of Sec.~\ref{sec:constraints}:
boundedness from below, perturbative unitarity, and an electroweak vacuum that
is the physical vacuum today. We then impose the relic-density criterion of
Sec.~\ref{subsec:relic_density_criterion}, followed by the direct-detection
condition in Eq.~\eqref{eq:dirdet} and, for $M_S<M_h/2$, the invisible
Higgs-decay bound.

This filtering procedure is also relevant for the finite-temperature analysis.
Direct-detection limits do not simply reduce the number of viable dark-matter
points; they determine whether the Higgs portal can remain large enough to
modify the electroweak thermal potential. The four regimes therefore differ not
only in their relic-production mechanisms, but also in how strongly present-day
dark-matter searches restrict the scalar dynamics relevant for the phase
transition.

\subsubsection{Scenario I: Two-Component Thermal Regime with \texorpdfstring{$M_\psi < M_S < 2M_\psi$}{Mpsi < MS < 2Mpsi}}

Scenario~I corresponds to the thermal two-component regime with
$M_\psi<M_S<2M_\psi$. Both particles are stable and thermally produced. Among
the viable points, the relic density is typically fermion dominated, often with
$\xi_\psi\gtrsim0.9$, in qualitative agreement with
Ref.~\cite{YagunaZapata2022}. This hierarchy in the relic fractions is driven
by the interaction structure: the scalar can be efficiently depleted by
Higgs-portal annihilation into SM states, whereas the fermion abundance is
controlled by conversion and semi-annihilation within the dark sector.

\begin{figure}[!htb] 
    \centering
    \begin{subfigure}{1.0\textwidth}
        \centering
        \begin{minipage}{0.48\textwidth} 
            \centering
            \includegraphics[width=\textwidth]{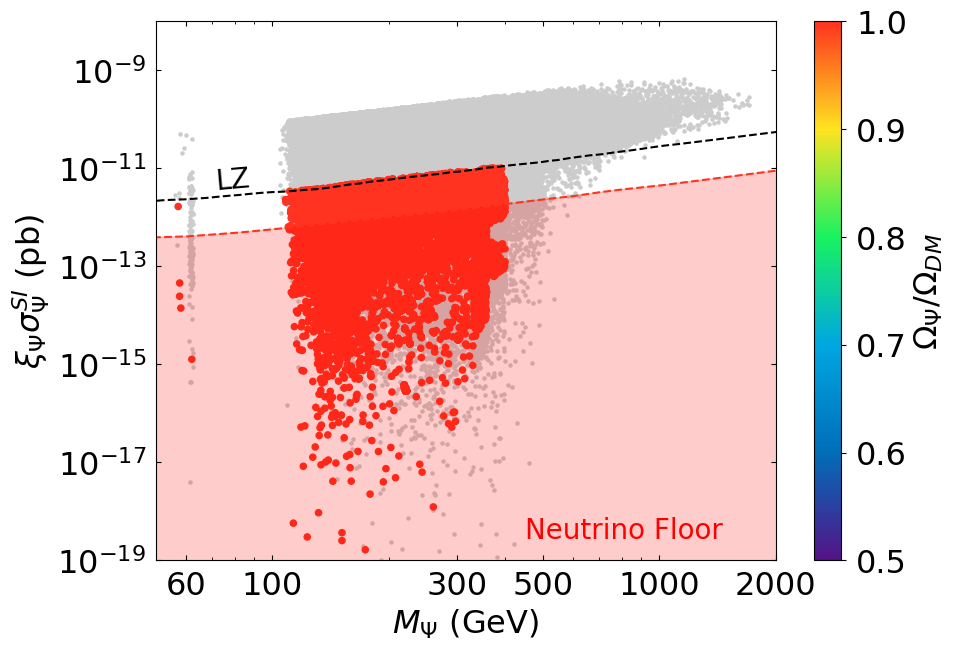}
        \end{minipage}
        \hfill
        \begin{minipage}{0.48\textwidth} 
            \centering
            \includegraphics[width=\textwidth]{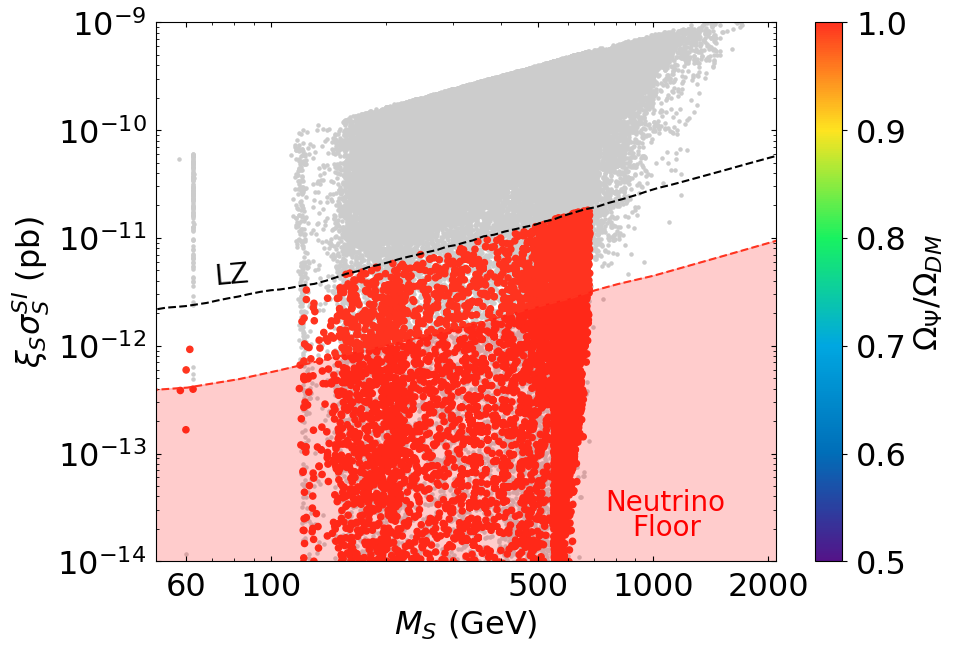}
        \end{minipage}
        \caption{\justifying 
        Direct-detection constraints for Scenario~I ($M_\psi<M_S<2M_\psi$). 
        All points satisfy the theoretical constraints and reproduce the observed relic abundance. 
        \textbf{Left:} Rescaled fermionic spin-independent cross section, $\xi_\psi \sigma_\psi^{\rm SI}$, as a function of $M_\psi$. 
        \textbf{Right:} Rescaled scalar spin-independent cross section, $\xi_S \sigma_S^{\rm SI}$, as a function of $M_S$. 
        Grey points are excluded by the latest LZ spin-independent bound through the combined direct-detection criterion in Eq.~\eqref{eq:dirdet}, while coloured points remain allowed. 
        The colour scale shows the fermionic relic fraction $\xi_\psi$.}
        \label{fig:Caso1a}
    \end{subfigure}

    \vspace{0.5em} 

    \begin{subfigure}{1.0\textwidth}
        \centering
        \begin{minipage}{0.48\textwidth} 
            \centering
            \includegraphics[width=\textwidth]{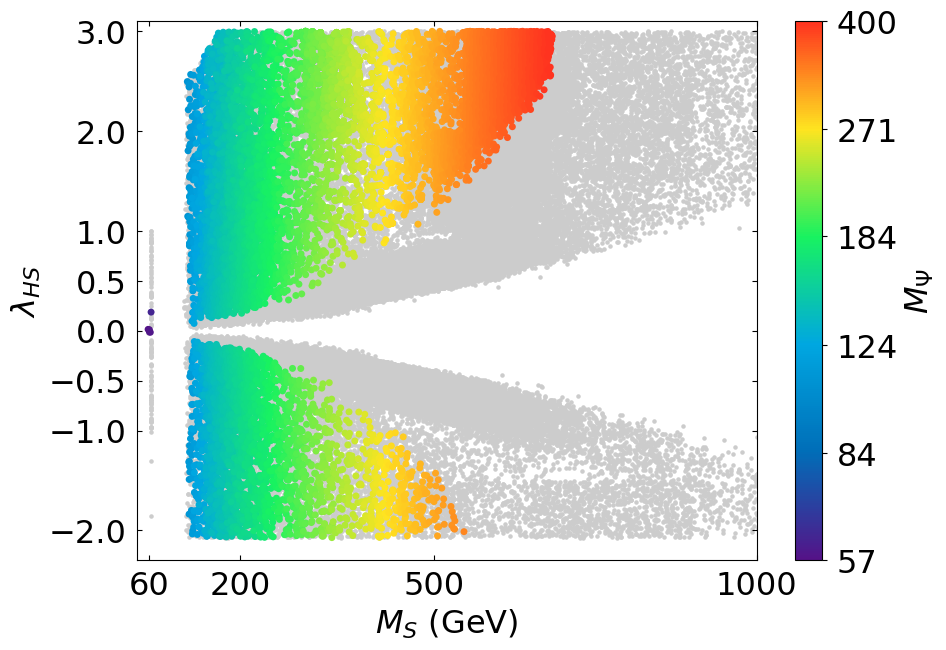}
        \end{minipage}
        \hfill
        \begin{minipage}{0.48\textwidth} 
            \centering
            \includegraphics[width=\textwidth]{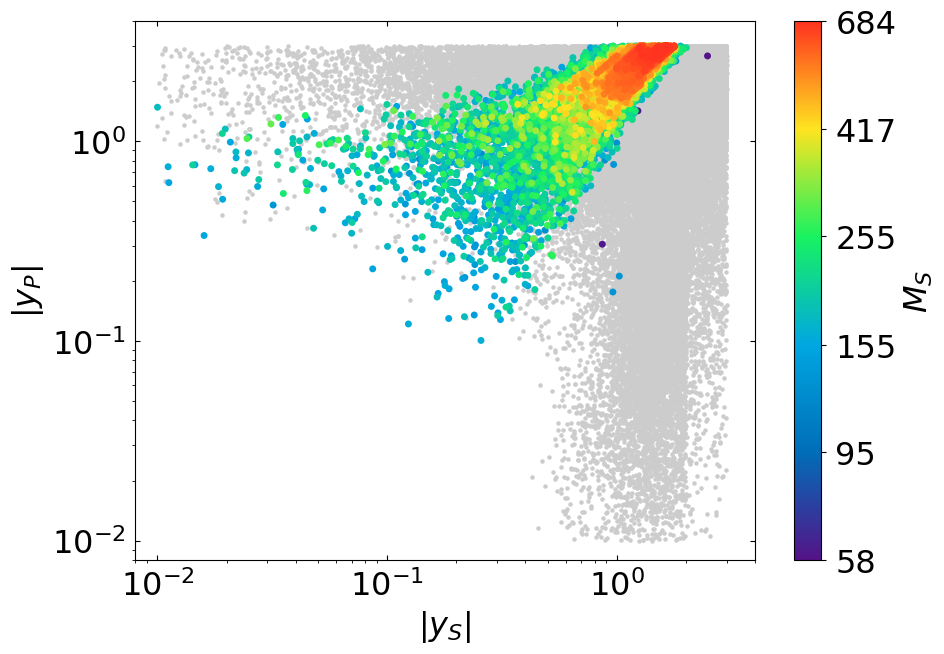}
        \end{minipage}
        \caption{\justifying 
        Parameter correlations for Scenario~I after imposing theoretical consistency, relic density, and direct-detection bounds. 
        \textbf{Left:} Higgs-portal coupling $\lambda_{HS}$ as a function of the scalar mass $M_S$, with the fermion mass shown in the colour scale. 
        \textbf{Right:} Yukawa couplings $y_s$ and $y_p$, with the scalar mass encoded in the colour scale.}
        \label{fig:Caso1b}
    \end{subfigure}
    \caption{\justifying 
    Viable parameter-space structure for Scenario~I, the thermal two-component regime with $M_\psi<M_S<2M_\psi$. The surviving region is typically dominated by the fermionic component, while direct-detection bounds remove a significant fraction of the relic-density-compatible parameter space.}
    \label{fig:Caso1}
\end{figure}

Figure~\ref{fig:Caso1a} shows that the latest LZ spin-independent bound removes
a significant part of the relic-density-compatible region. However, unlike in
the scalar-dominated case, the surviving points are not confined to the Higgs
resonance region. The allowed dark-sector masses extend approximately from
$100~\mathrm{GeV}$ to $1~\mathrm{TeV}$ in the scan. The exclusion is determined
by the combined rate criterion in Eq.~\eqref{eq:dirdet}: a point is removed when
the total predicted recoil rate, after relic-fraction rescaling, exceeds the
single-component LZ limit.

The surviving parameter correlations are shown in Fig.~\ref{fig:Caso1b}. The
Higgs-portal coupling can still reach sizeable positive values after all
dark-matter constraints are imposed. This is important for the subsequent
finite-temperature analysis, because the same portal coupling controls the
communication between the Higgs and singlet directions in the effective
potential.

\subsubsection{Scenario II: Two-Component Thermal Regime with \texorpdfstring{$M_\psi > M_S$}{Mpsi > MS}}

Scenario~II is the complementary thermal two-component regime, with
$M_S<M_\psi$. Both particles remain stable, but the lighter scalar couples to
nucleons at tree level through Higgs exchange. As a result, the scalar component
is more strongly constrained by direct detection than in Scenario~I.

\begin{figure}[!htb] 
    \centering
    \begin{subfigure}{1.0\textwidth}
        \centering
        \begin{minipage}{0.48\textwidth}
            \centering
            \includegraphics[width=\textwidth]{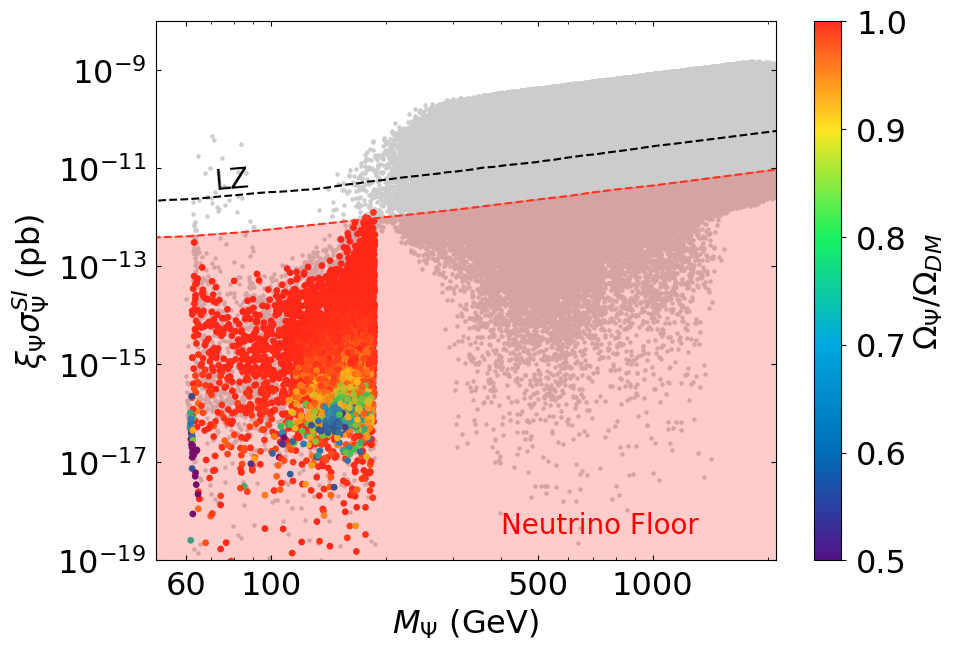}
        \end{minipage}
        \hfill
        \begin{minipage}{0.48\textwidth}
            \centering
            \includegraphics[width=\textwidth]{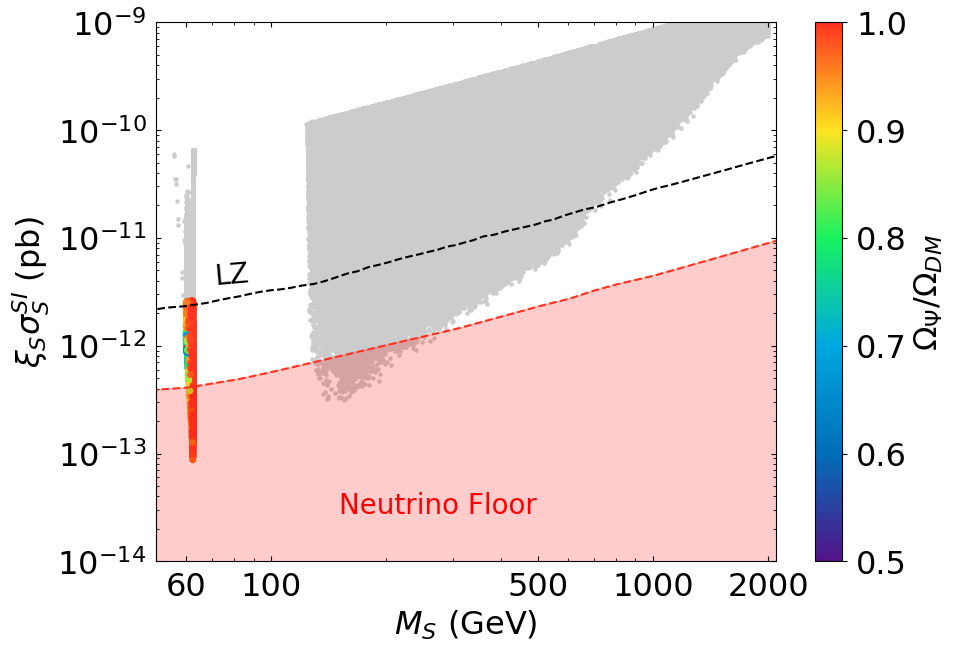}
        \end{minipage}
        \caption{\justifying 
        Direct-detection constraints for Scenario~II ($M_S<M_\psi$). 
        All points satisfy the theoretical constraints and reproduce the observed relic abundance. 
        \textbf{Left:} Rescaled fermionic spin-independent cross section, $\xi_\psi \sigma_\psi^{\rm SI}$, as a function of $M_\psi$. 
        \textbf{Right:} Rescaled scalar spin-independent cross section, $\xi_S \sigma_S^{\rm SI}$, as a function of $M_S$. 
        Grey points are excluded by the latest LZ spin-independent bound through the combined direct-detection criterion in Eq.~\eqref{eq:dirdet}, while coloured points remain allowed. 
        The colour scale shows the fermionic relic fraction $\xi_\psi$.}
        \label{fig:Caso2a}
    \end{subfigure}

    \vspace{0.5em} 

    \begin{subfigure}{1.0\textwidth}
        \centering
        \begin{minipage}{0.48\textwidth}
            \centering
            \includegraphics[width=\textwidth]{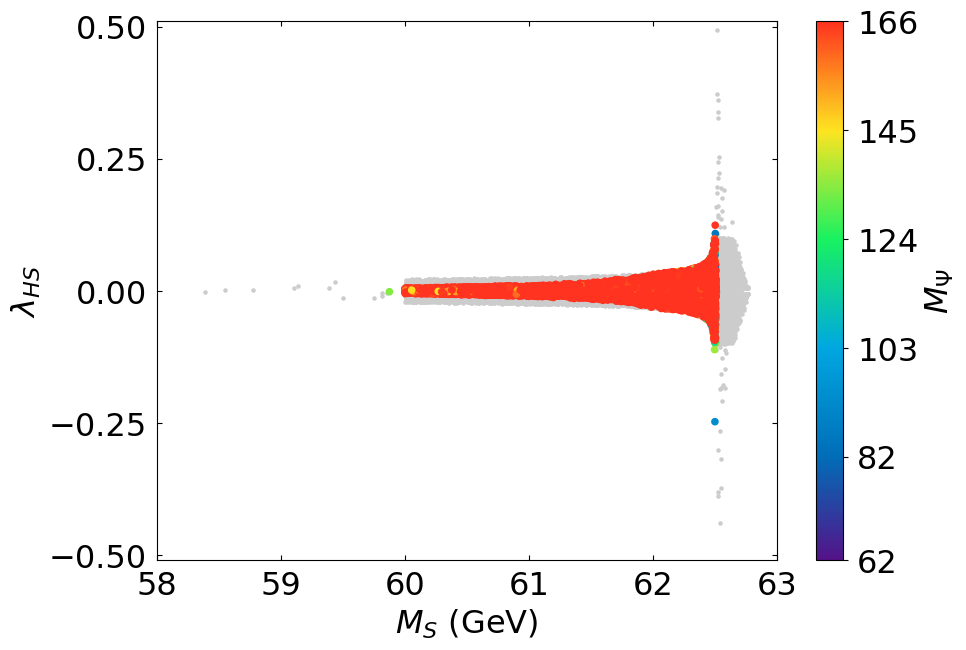}
        \end{minipage}
        \hfill
        \begin{minipage}{0.48\textwidth} 
            \centering
            \includegraphics[width=\textwidth]{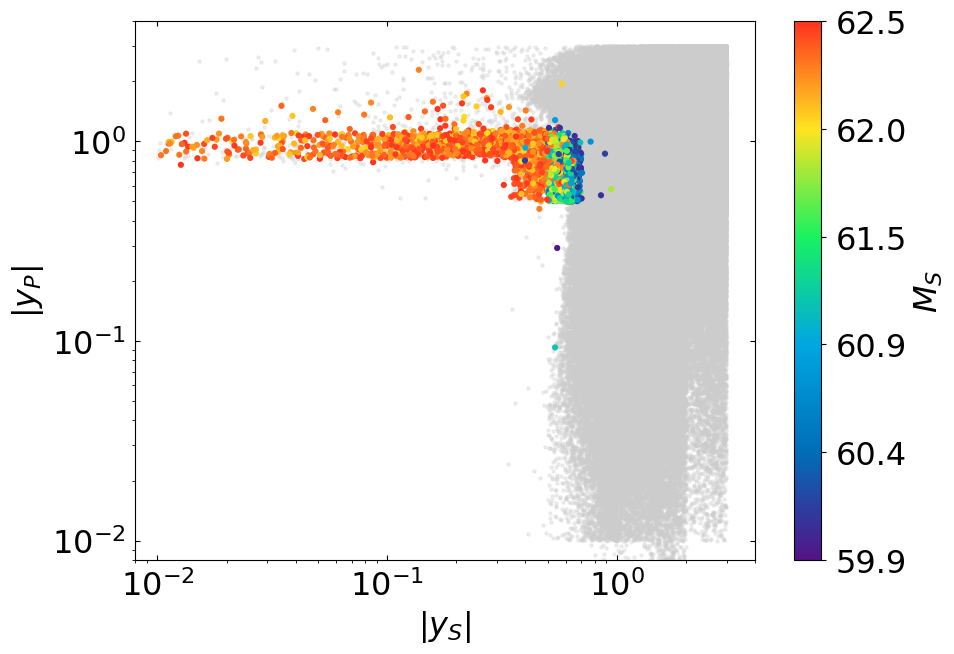}
        \end{minipage}
        \caption{\justifying 
        Parameter correlations for Scenario~II after imposing theoretical consistency, relic density, and direct-detection bounds. 
        \textbf{Left:} Higgs-portal coupling $\lambda_{HS}$ as a function of the scalar mass $M_S$, with the fermion mass shown in the colour scale. 
        \textbf{Right:} Yukawa couplings $y_s$ and $y_p$, with the scalar mass encoded in the colour scale.}
        \label{fig:Caso2b}
    \end{subfigure}
    \caption{\justifying 
    Viable parameter-space structure for Scenario~II, the thermal two-component regime with $M_S<M_\psi$. Direct-detection bounds strongly restrict the scalar component and drive the surviving points towards the Higgs-resonance region.}
    \label{fig:Caso2}
\end{figure}

As shown in Fig.~\ref{fig:Caso2a}, the surviving region is concentrated near
the Higgs resonance, $M_S\simeq M_h/2$. In this region, resonant annihilation
can deplete the scalar abundance efficiently even for small Higgs-portal
couplings. Away from the resonance, small portal couplings tend to leave an
excessive scalar relic abundance, while larger portal couplings typically
produce a direct-detection rate incompatible with the latest LZ
spin-independent bound. The allowed region is therefore shaped by the tension
between scalar freeze-out and direct detection.

This behaviour is reflected in Fig.~\ref{fig:Caso2b}. The Higgs-portal coupling
is driven to small values, while the Yukawa couplings control the conversion
and semi-annihilation processes that shape the fermionic abundance. As in
Scenario~I, the viable points are typically fermion dominated, indicating that
fermion dominance is a common feature of the thermal two-component regions
explored here.

\subsubsection{Scenario III: Mixed Two-Component WIMP--FIMP Regime with \texorpdfstring{$M_S < 2M_\psi$}{MS < 2Mpsi}}

Scenario~III combines a thermal scalar WIMP with a feebly coupled fermionic
FIMP. Since $M_S<2M_\psi$, the decay $S\to\psi\psi$ is kinematically forbidden,
and both particles remain stable today. The present relic density is therefore
shared between a thermal scalar component and a non-thermal fermionic component.

In this regime, the Higgs portal controls the scalar freeze-out abundance,
while the tiny Yukawa couplings determine the out-of-equilibrium production of
$\psi$. Increasing $|\lambda_{HS}|$ enhances the scalar annihilation rate into
SM states and suppresses the scalar relic fraction, thereby increasing the
relative weight of the fermionic component in the total abundance. Conversely,
smaller portal couplings allow the scalar to contribute more substantially.
Because the fermion is feebly coupled, its loop-induced direct-detection rate is
negligible, and present direct-detection limits mainly constrain the stable
scalar component.

\begin{figure}[!ht]
    \centering
    \begin{minipage}{0.49\textwidth}
        \centering
        \includegraphics[width=\textwidth]{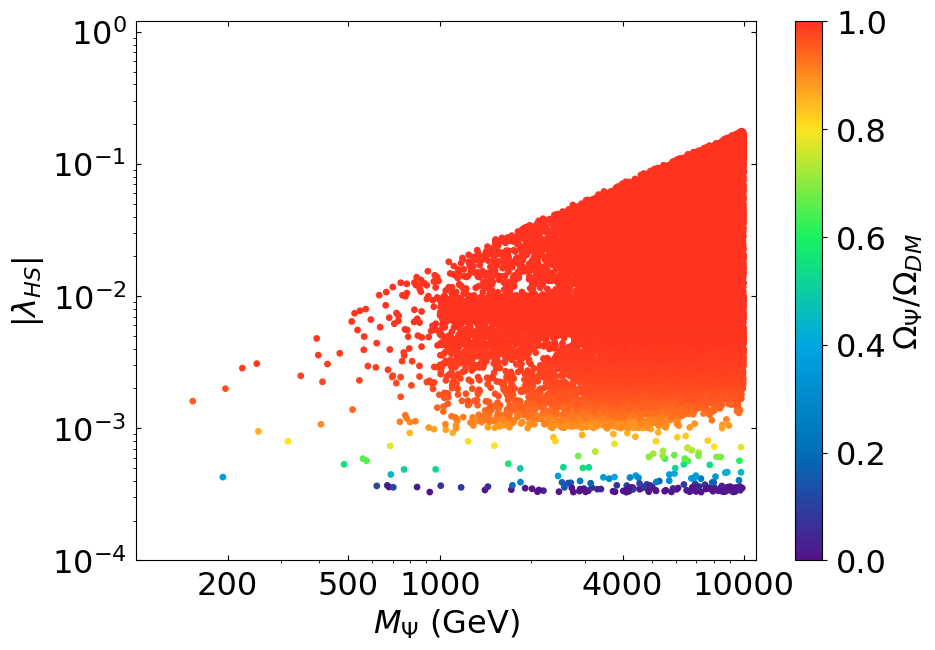}
    \end{minipage}
    \caption{\justifying 
    Viable parameter space for Scenario~III, the mixed two-component WIMP--FIMP regime with $M_S<2M_\psi$. The Higgs-portal coupling $\lambda_{HS}$ is shown against the fermion mass $M_\psi$, with the fermionic relic fraction $\Omega_\psi/\Omega_{\rm DM}$ encoded in the colour scale. The surviving points reflect the interplay between scalar freeze-out, fermion freeze-in, and the latest LZ spin-independent bound on the stable scalar component.}
    \label{fig:Caso3}
\end{figure}

As shown in Fig.~\ref{fig:Caso3}, viable points occur in regions where the
stable scalar component remains compatible with direct detection after
relic-fraction rescaling. This typically pushes the scalar sector towards the
Higgs-resonance region, $M_S\simeq M_h/2$, where resonant annihilation can
reduce the scalar abundance without requiring a large portal coupling. Away
from this region, the scalar relic fraction and the Higgs-mediated scattering
rate are difficult to reconcile with the latest LZ spin-independent bound.

Thus, Scenario~III is characterised by a stable scalar component whose portal
interaction must simultaneously account for scalar freeze-out and remain
compatible with direct detection. The allowed region is therefore shaped by the
same tension already present in scalar Higgs-portal dark matter, with the
additional possibility that a feebly coupled fermion contributes a non-thermal
fraction of the total relic abundance.

\subsubsection{Scenario IV: Effectively One-Component WIMP--FIMP Regime with Scalar Decay}

Scenario~IV corresponds to the decay-driven regime with $M_S>2M_\psi$. The
scalar is thermally produced through the Higgs portal and subsequently decays
into fermions, leaving a purely fermionic dark-matter abundance at late times.
For the small Yukawa couplings considered in this regime, the fermion remains
out of equilibrium, and its final abundance receives contributions from
freeze-in production and from the decay of the thermal scalar population.

This regime differs qualitatively from the previous two-component cases. Since
the scalar does not survive as a present-day dark-matter component, the
tree-level scalar-nucleon bound does not apply to the late-time relic
abundance. The surviving fermion is feebly coupled, and its loop-induced
scattering rate is negligible in the parameter region considered. Consequently,
the Higgs-portal coupling is not forced into the small-coupling or
Higgs-resonance regions.

\begin{figure}[!ht]
    \centering
    \begin{minipage}{0.49\textwidth}
        \centering
        \includegraphics[width=\textwidth]{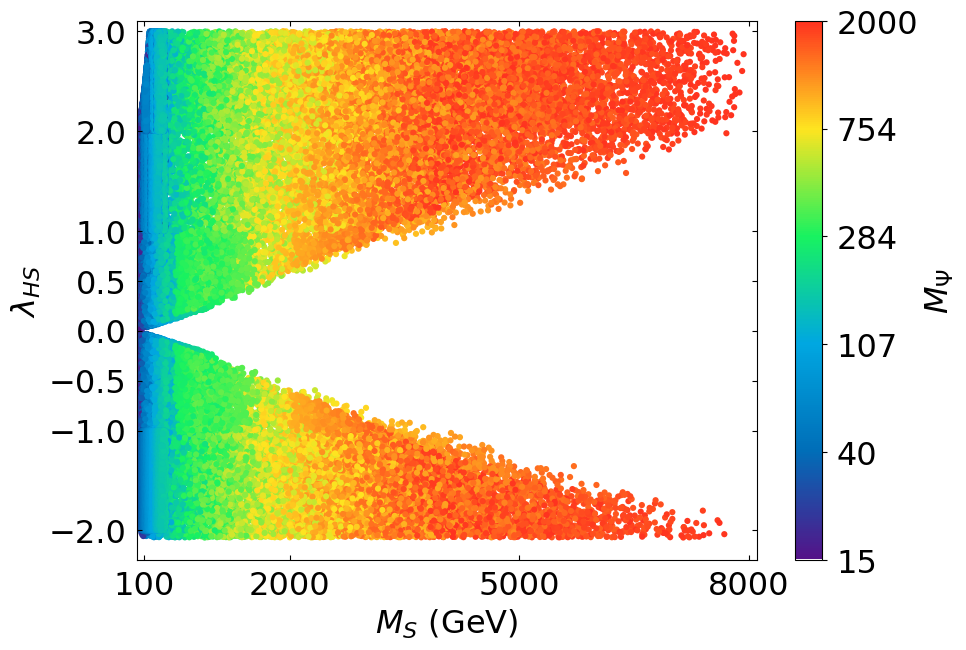}
    \end{minipage}
    \caption{\justifying 
    Viable parameter space for Scenario~IV, the effectively one-component
    WIMP--FIMP regime with $M_S>2M_\psi$. The Higgs-portal coupling
    $\lambda_{HS}$ is shown against the scalar mass $M_S$, with the fermion
    mass $M_\psi$ encoded in the colour scale. Since the scalar decays before
    today, direct-detection bounds on scalar dark matter do not constrain the
    present-day relic abundance. Sizeable portal couplings therefore remain
    allowed, subject to theoretical consistency and the relic-density
    requirement.}
    \label{fig:Caso4}
\end{figure}

As shown in Fig.~\ref{fig:Caso4}, sizeable values of $\lambda_{HS}$ can remain
compatible with all imposed dark-matter constraints. Although the scalar is
absent from the present-day dark-matter abundance, it was thermally populated in
the early Universe through the Higgs portal. Scenario~IV therefore provides a
case in which the scalar sector remains active in the early thermal history
without being directly constrained as a stable halo component.

Combining the four regimes, we find that the viable dark-matter regions are
often fermion dominated, but for different reasons. In the thermal
two-component regimes, the scalar can be efficiently depleted through
Higgs-portal annihilation. In Scenario~III, the stable scalar component is
strongly constrained by direct detection and is typically pushed towards the
Higgs-resonance region. In Scenario~IV, the scalar decays into fermions and is
absent from the present-day halo, so direct detection constrains only the
feebly coupled fermionic relic. Thus, the phenomenological role of the Higgs
portal depends crucially on whether the scalar survives until today.

\section{Thermal Effective Potential and Finite-Temperature Framework}
\label{sec:effective_potential}

To analyse the electroweak phase transition, we use the one-loop effective
potential at zero and finite temperature in a fixed perturbative prescription.
The Coleman--Weinberg contribution is evaluated in the
$\overline{\mathrm{MS}}$ scheme, and no additional finite counterterms are
introduced to enforce the tree-level vacuum conditions or scalar masses at one
loop. The zero-temperature vacuum structure is therefore imposed at tree level,
while one-loop corrections are included according to this fixed prescription.
Unless otherwise stated, all one-loop quantities are computed in Landau gauge.
Consequently, the transition parameters extracted from this potential should be
understood within this gauge and renormalisation prescription, with the residual
scale and scheme dependence inherent to perturbative finite-temperature
calculations.

\subsection{Background fields and field-dependent masses}

We introduce the classical background fields $(\phi,\sigma)$ through
\begin{equation}
H=
\begin{pmatrix}
G^+\\[2mm]
\dfrac{\phi+h+iG^0}{\sqrt{2}}
\end{pmatrix},
\qquad
S=\sigma+s,
\label{eq:background_fields}
\end{equation}
where $\phi$ and $\sigma$ denote the Higgs and singlet background fields,
respectively. The tree-level potential in these directions is
\begin{equation}
V_0(\phi,\sigma)
=
-\frac{1}{2}\mu_H^2\phi^2
+\frac{1}{4}\lambda_H\phi^4
-\frac{1}{2}\mu_S^2 \sigma^2
+\frac{1}{4}\lambda_S \sigma^4
+\frac{1}{4}\lambda_{HS}\phi^2 \sigma^2.
\label{eq:tree_background}
\end{equation}

The scalar field-dependent mass matrix is obtained from the Hessian of
$V_0(\phi,\sigma)$ in the $(h,s)$ directions,
\begin{equation}
\mathcal{M}_{\mathrm{scal}}^2(\phi,\sigma)
=
\begin{pmatrix}
\mathcal{M}_{11}^2 & \mathcal{M}_{12}^2\\
\mathcal{M}_{12}^2 & \mathcal{M}_{22}^2
\end{pmatrix},
\label{eq:scalar_mass_matrix}
\end{equation}
with
\begin{align}
\mathcal{M}_{11}^2(\phi,\sigma)
&=
-\mu_H^2 + 3\lambda_H\phi^2 + \frac{1}{2}\lambda_{HS}\sigma^2,
\\
\mathcal{M}_{22}^2(\phi,\sigma)
&=
-\mu_S^2 + 3\lambda_S \sigma^2 + \frac{1}{2}\lambda_{HS}\phi^2,
\\
\mathcal{M}_{12}^2(\phi,\sigma)
&=
\lambda_{HS}\phi \sigma.
\end{align}
Its eigenvalues are
\begin{equation}
m_{h_{1,2}}^2(\phi,\sigma)
=
\frac{1}{2}
\left[
\mathcal{M}_{11}^2+\mathcal{M}_{22}^2
\pm
\sqrt{
\left(\mathcal{M}_{11}^2-\mathcal{M}_{22}^2\right)^2
+
4\left(\mathcal{M}_{12}^2\right)^2
}
\right].
\label{eq:scalar_eigenvalues}
\end{equation}
The Goldstone modes have the common field-dependent mass
\begin{equation}
m_G^2(\phi,\sigma)
=
-\mu_H^2+\lambda_H\phi^2+\frac{1}{2}\lambda_{HS}\sigma^2.
\label{eq:goldstone_mass}
\end{equation}

The gauge-boson and top-quark masses retain their standard dependence on the
Higgs background,
\begin{equation}
m_W^2(\phi)=\frac{g^2}{4}\phi^2,
\qquad
m_Z^2(\phi)=\frac{g^2+g'^2}{4}\phi^2,
\qquad
m_t^2(\phi)=\frac{y_t^2}{2}\phi^2.
\label{eq:gauge_top_masses}
\end{equation}

In the fermion sector, a non-vanishing singlet background induces
Majorana-like entries. Writing the fields in the basis
\begin{equation}
\Psi_R \equiv
\begin{pmatrix}
\psi_L^c\\
\psi_R
\end{pmatrix},
\qquad
\overline{\Psi}_L=
\begin{pmatrix}
\overline{\psi_L} & \overline{\psi_R^c}
\end{pmatrix},
\end{equation}
the mass term takes the form
\begin{equation}
\mathcal{L}_{\rm mass}^{(\psi)}
=
-\frac{1}{2}\overline{\Psi}_L\,\mathcal{M}_\psi(\sigma)\,\Psi_R
+\text{h.c.},
\end{equation}
with
\begin{equation}
\mathcal{M}_\psi(\sigma)
=
\begin{pmatrix}
(y_s-y_p)\sigma & M_\psi \\[2mm]
M_\psi & (y_s+y_p)\sigma
\end{pmatrix}.
\label{eq:fermion_mass_matrix}
\end{equation}
The corresponding field-dependent squared masses are
\begin{equation}
m_{\psi_{1,2}}^2(\sigma)
=
\left(
y_s \sigma \pm \sqrt{M_\psi^2+y_p^2 \sigma^2}
\right)^2.
\label{eq:fermion_field_masses}
\end{equation}
At $\sigma=0$, the physical masses are degenerate,
$m_{\psi_1}(0)=m_{\psi_2}(0)=M_\psi$, recovering the Dirac fermion of the
$\mathbb{Z}_4$-preserving vacuum.

\subsection{One-loop effective potential in \texorpdfstring{$\overline{\mathrm{MS}}$}{MSbar}}

For convenience, we define
\begin{equation}
f(x;C)\equiv x^2\left[\ln\!\left(\frac{x}{\mu_R^2}\right)-C\right],
\label{eq:aux_function_f}
\end{equation}
where $x$ has mass dimension two. The zero-temperature one-loop correction is
written as
\begin{equation}
V_{CW}^{\overline{\mathrm{MS}}}(\phi,\sigma)
=
\frac{1}{64\pi^2}
\sum_i
(-1)^{F_i} n_i\,
f\!\left(m_i^2(\phi,\sigma);C_i\right),
\label{eq:VCW_MSbar}
\end{equation}
where $F_i=1$ for fermions and $F_i=0$ for bosons. The constants $C_i$ take
their standard $\overline{\mathrm{MS}}$ values,
\begin{equation}
C_i=
\begin{cases}
3/2, & \text{for scalars and fermions},\\
5/6, & \text{for gauge bosons}.
\end{cases}
\end{equation}
When field-dependent scalar squared masses become negative away from a local
minimum, we use the real part of the Coleman--Weinberg potential, equivalently
replacing the logarithm by $\ln|m_i^2|$ in Eq.~\eqref{eq:VCW_MSbar}.

The sum includes the scalar eigenstates $h_{1,2}$, the Goldstone modes, the
electroweak gauge bosons, the top quark, and the two singlet-fermion
eigenstates $\psi_{1,2}$. The corresponding degrees of freedom are
\begin{equation}
n_{h_1}=n_{h_2}=1,\quad
n_G=3,\quad
n_W=6,\quad
n_Z=3,\quad
n_t=12,\quad
n_{\psi_1}=n_{\psi_2}=2.
\label{eq:CW_degrees_of_freedom}
\end{equation}
Massless field-independent contributions are omitted.

We fix the renormalisation scale to
\begin{equation}
\mu_R=\frac{m_t}{2}.
\end{equation}
At fixed order, this scale choice should be regarded as part of the definition
of the effective potential used in the scan. The Coleman--Weinberg contribution
retains a residual renormalisation-scale dependence, and different
renormalisation prescriptions can shift the numerical values of the transition
parameters. We adopt $\mu_R=m_t/2$ following related singlet-extension analyses
in which this choice was found to give close agreement between
$\overline{\mathrm{MS}}$ and OS-like treatments
\cite{Oikonomou:2024ewpt,Chiang:2018gsn,Athron:2022qpo}. Broader discussions of
renormalisation-scale dependence and fixed-order limitations in cosmological
phase transitions can be found in Refs.~\cite{Croon:2020cgk,Gould:2021oba}. A
systematic reduction of this residual dependence, for instance through RGE
improvement or an on-shell counterterm prescription, is beyond the scope of the
present work.

The same field-dependent spectrum is used in the finite-temperature
contribution below, with the bosonic modes resummed as described in the next
subsection.

\subsection{Finite-temperature contributions and Parwani resummation}

The unresummed finite-temperature contribution is
\begin{equation}
V_T^{(0)}(\phi,\sigma,T)
=
\frac{T^4}{2\pi^2}
\left[
\sum_{i\in B} n_i\,
J_B\!\left(\frac{m_i^2(\phi,\sigma)}{T^2}\right)
-
\sum_{j\in F} n_j\,
J_F\!\left(\frac{m_j^2(\phi,\sigma)}{T^2}\right)
\right],
\label{eq:VT_unresummed}
\end{equation}
where the thermal functions are defined by
\begin{equation}
J_B(u)=
\int_0^\infty dq\,q^2
\ln\!\left(1-e^{-\sqrt{q^2+u}}\right),
\qquad
J_F(u)=
\int_0^\infty dq\,q^2
\ln\!\left(1+e^{-\sqrt{q^2+u}}\right).
\label{eq:thermal_functions}
\end{equation}
Here, $u$ is a dimensionless argument. The sets $B$ and $F$ denote bosonic and
fermionic modes, respectively, and the fermionic degrees of freedom are taken to be positive; the relative minus sign in Eq.~\eqref{eq:VT_unresummed} accounts for
fermionic statistics.

At high temperature, infrared-enhanced bosonic zero modes require daisy
resummation. We adopt the Parwani prescription~\cite{Parwani1992,Senaha2020,Bittar2025},
in which bosonic field-dependent masses are replaced by thermally corrected
masses,
\begin{equation}
m_i^2(\phi,\sigma)
\;\longrightarrow\;
\mathcal{M}_i^2(\phi,\sigma,T)
=
m_i^2(\phi,\sigma)+\Pi_i(T),
\label{eq:Parwani_replacement}
\end{equation}
with $\Pi_i(T)$ the corresponding thermal self-energies. This replacement is
applied to scalar modes and to the longitudinal gauge modes. Transverse gauge
modes and fermionic modes are not thermally resummed.

The Parwani prescription defines the resummation scheme used throughout the
numerical analysis. Other daisy-resummation prescriptions, such as
Arnold--Espinosa implementations, can lead to quantitative shifts in the
transition parameters. The results below should therefore be interpreted within
the resummation scheme specified here.

In the scalar sector, the leading thermal self-energies are diagonal in the
$(\phi,\sigma)$ basis. The off-diagonal entry is therefore still determined by
the tree-level mixing term,
$\mathcal{M}_{12}^2(\phi,\sigma)=\lambda_{HS}\phi\sigma$. The thermally
corrected scalar matrix is
\begin{equation}
\mathcal{M}_{\mathrm{scal},T}^2(\phi,\sigma,T)
=
\mathcal{M}_{\mathrm{scal}}^2(\phi,\sigma)
+
\Pi_{\mathrm{scal}}(T),
\label{eq:scalar_thermal_matrix}
\end{equation}
and its eigenvalues define the resummed scalar masses entering the effective
potential. An analogous diagonalisation is performed in the longitudinal
neutral gauge sector. The explicit thermal self-energies and resummed masses
used in the numerical analysis are collected in
Appendix~\ref{app:thermal_masses}.

With Parwani resummation, daisy effects are incorporated directly through the
mass replacement in Eq.~\eqref{eq:Parwani_replacement}; no separate ring term is
added. The effective potential is written as
\begin{equation}
V_{\mathrm{eff}}(\phi,\sigma,T)
=
V_0(\phi,\sigma)
+
V_{CW}^{\mathrm{P}}(\phi,\sigma,T)
+
V_T^{\mathrm{P}}(\phi,\sigma,T),
\label{eq:Veff_final_decomp}
\end{equation}
with
\begin{equation}
V_{CW}^{\mathrm{P}}(\phi,\sigma,T)
=
\frac{1}{64\pi^2}
\left[
\sum_{i\in B} n_i\,
f\!\left(\mathcal{M}_i^2(\phi,\sigma,T);C_i\right)
-
\sum_{j\in F} n_j\,
f\!\left(m_j^2(\phi,\sigma);C_j\right)
\right],
\label{eq:VCW_Parwani}
\end{equation}
and
\begin{equation}
V_T^{\mathrm{P}}(\phi,\sigma,T)
=
\frac{T^4}{2\pi^2}
\left[
\sum_{i\in B} n_i\,
J_B\!\left(\frac{\mathcal{M}_i^2(\phi,\sigma,T)}{T^2}\right)
-
\sum_{j\in F} n_j\,
J_F\!\left(\frac{m_j^2(\phi,\sigma)}{T^2}\right)
\right].
\label{eq:VT_Parwani}
\end{equation}
In these expressions, the bosonic sums include the scalar modes, the Goldstone
modes, and the electroweak gauge degrees of freedom with the resummation
specified above. The fermionic sums include the top quark and the two
singlet-fermion eigenstates. In the FIMP regimes, the field dependence of the
singlet-fermion thermal contribution is suppressed by the feeble Yukawa
couplings; keeping these modes in the sums provides a uniform treatment across
the four scenarios without affecting the scalar thermal dynamics.

As in other perturbative finite-temperature analyses, local imaginary parts may
appear in regions where some bosonic squared masses become negative. In the
numerical analysis, we use the real part of the effective potential when
tracking the relevant minima and their evolution with temperature. The
associated imaginary parts are treated as a limitation of the local
perturbative expansion, not as physical instabilities of the phases being
compared.

The resulting $V_{\mathrm{eff}}(\phi,\sigma,T)$ is the potential used in
Sec.~\ref{sec:EWPT} to determine the phase structure, identify critical
temperatures, and select the benchmark points for the subsequent nucleation and
gravitational-wave analysis.

\section{Electroweak Phase Transition}
\label{sec:EWPT}

We now use the finite-temperature effective potential constructed in
Sec.~\ref{sec:effective_potential} to study the electroweak phase transition
in the dark-matter viable regions identified in Sec.~\ref{sec:dark_matter}.
The phase-transition analysis is therefore not an independent scan of the
scalar potential, but a second step applied to points that already satisfy the
zero-temperature and dark-matter constraints. For each viable configuration, we
determine whether the finite-temperature potential supports a first-order
electroweak transition satisfying the strong-transition criterion. Bubble
nucleation and the connection to gravitational-wave production are treated
separately in Sec.~\ref{sec:GW}.

At high temperature, thermal corrections tend to restore the symmetries of the
scalar potential. As the Universe cools, new minima may appear away from the
origin, and the global minimum can change discontinuously. In the present
model, the transition may proceed directly towards the electroweak vacuum or
through an intermediate minimum along the singlet direction. A relevant
two-step pattern is
\begin{equation}
(0,0)
\;\xrightarrow{\ T_1 \ }\;
(0,w)
\;\xrightarrow{\ T_2 \ }\;
(v',0),
\label{eq:two_step_transition}
\end{equation}
where the first step breaks the dark symmetry along the singlet direction, leaving a residual $\mathbb{Z}_2$ at temperature $T_1$,
while the second step restores the $\mathbb{Z}_4$ symmetry and breaks the
electroweak symmetry at $T_2<T_1$. The final state is therefore the present-day vacuum
selected in Sec.~\ref{subsec:vacuum_structure}, with the dark symmetry
preserved.

The critical temperature $T_c$ is defined by the degeneracy condition
\begin{equation}
V_{\rm eff}\!\left(\boldsymbol{\Phi}^{\rm high},T_c\right)
=
V_{\rm eff}\!\left(\boldsymbol{\Phi}^{\rm low},T_c\right),
\label{eq:critical_temperature_condition}
\end{equation}
where
\begin{equation}
\boldsymbol{\Phi}=(\Phi_1,\Phi_2)=(\phi,\sigma),
\qquad
\Delta\Phi_i\equiv
\Phi_i^{\rm low}-\Phi_i^{\rm high}.
\end{equation}
For the electroweak-breaking step in Eq.~\eqref{eq:two_step_transition},
\begin{equation}
\boldsymbol{\Phi}^{\rm high}=(0,w_c),
\qquad
\boldsymbol{\Phi}^{\rm low}=(v_c,0),
\end{equation}
with both field values evaluated at $T_c$.

The total discontinuity in field space is measured by
\begin{equation}
\gamma_\Phi
\equiv
\frac{\sqrt{\Delta\Phi_i\Delta\Phi_i}}{T_c},
\label{eq:gamma_field_space}
\end{equation}
where repeated field-space indices are summed over. For electroweak sphaleron
suppression, however, the relevant order parameter is the projection of this
jump onto the Higgs direction, since only $\phi$ breaks
$SU(2)_L\times U(1)_Y$. We therefore define
\begin{equation}
\zeta_c
\equiv
\frac{\sqrt{\Delta\Phi_i P^{\rm EW}_{ij}\Delta\Phi_j}}{T_c},
\qquad
P^{\rm EW}_{ij}=
\begin{pmatrix}
1 & 0\\
0 & 0
\end{pmatrix}.
\label{eq:zeta_c_projected}
\end{equation}
For the transition $(0,w_c)\to(v_c,0)$, this reduces to
\begin{equation}
\zeta_c=\frac{v_c}{T_c}.
\end{equation}
We use
\begin{equation}
\zeta_c\gtrsim 1
\label{eq:sfoewpt_condition}
\end{equation}
as the working criterion for a strong first-order electroweak transition at
the critical temperature. This criterion should be understood as an
order-parameter test within the finite-temperature effective-potential
framework, not as a full computation of baryon-number preservation. A more
precise treatment would require the sphaleron energy in the finite-temperature
two-field background and is gauge and model dependent~\cite{patel2011baryon}.
The criterion adopted here follows the standard practice in singlet extensions
of the SM~\cite{amine2007,profumo2007singlet}.

\subsection{Numerical implementation}
\label{subsec:ewpt_numerics}

The phase structure is analysed numerically with
\texttt{PhaseTracer2}~\cite{athron2025phasetracer2}. For each point that
satisfies the theoretical and dark-matter constraints, we follow the
temperature evolution of the local minima of
$V_{\rm eff}(\phi,\sigma,T)$ in the two-field space
$\boldsymbol{\Phi}=(\phi,\sigma)$.

The analysis is performed separately for the four dark-matter regimes defined
in Sec.~\ref{subsec:dark_matterProduction}. Each point passed to the
finite-temperature analysis includes a value of $\lambda_S$ assigned through
the vacuum-preserving prescription of Sec.~\ref{subsec:vacuum_structure},
namely $\lambda_S=\lambda_S^{\rm base}+a$, with $a$ sampled in the
point-dependent theoretically allowed interval of Eq.~\eqref{eq:a_scan_interval}.
This ensures that the scalar self-coupling used in the thermal potential is
consistent with boundedness from below, perturbative unitarity,
and the zero-temperature vacuum requirement.

For each thermal history, we identify the critical temperature $T_c$ of any
first-order transition involving the electroweak-breaking phase and determine
the corresponding pair of degenerate minima. We then compute the projected
order parameter $\zeta_c$ defined in Eq.~\eqref{eq:zeta_c_projected}.
Configurations with $\zeta_c\gtrsim 1$ are classified as satisfying the
strong-transition criterion at the critical temperature.

\subsection{Phase-transition patterns in the viable dark-matter regions}
\label{subsec:ewpt_results}

\begin{figure}[!htb]
    \centering

    \begin{minipage}{0.47\textwidth}
        \centering
        \includegraphics[width=\textwidth]{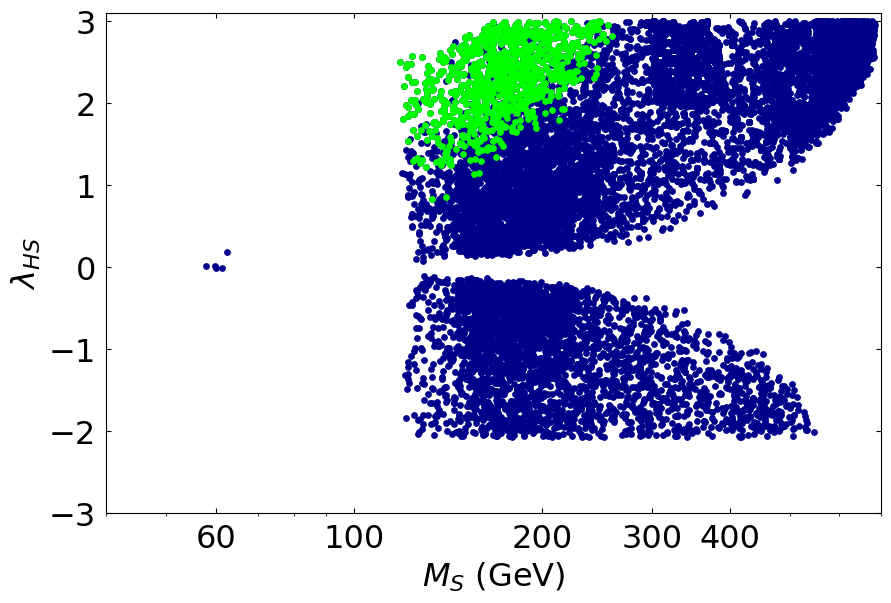}
    \end{minipage}\hfill
    \begin{minipage}{0.47\textwidth}
        \centering
        \includegraphics[width=\textwidth]{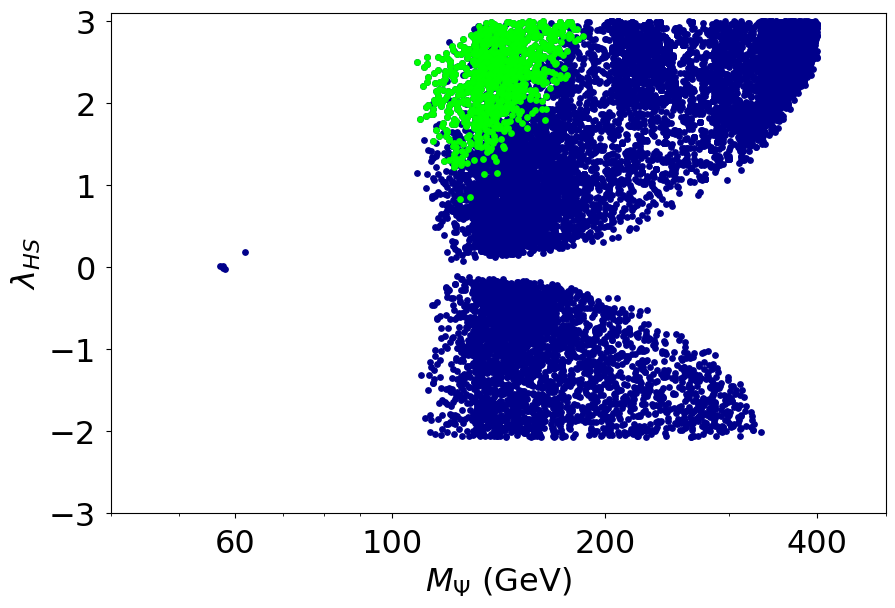}
    \end{minipage}
    
    \vspace{0.25cm}

    \begin{minipage}{0.47\textwidth}
        \centering
        \includegraphics[width=\textwidth]{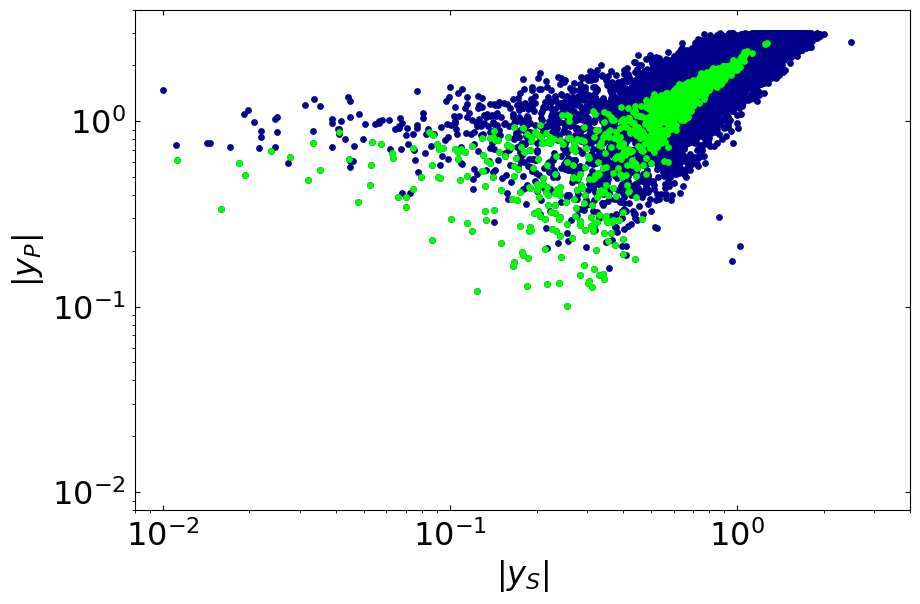}
    \end{minipage}

\caption{
Correlations between the projected order parameter
$\zeta_c=v_c/T_c$ and the model parameters in Scenario~I, corresponding to the
thermal two-component regime $M_\psi<M_S<2M_\psi$. Green points yield a strong
electroweak phase transition, $\zeta_c\gtrsim 1$, whereas blue points fall below
this threshold. Only configurations satisfying the theoretical and dark-matter
constraints discussed in the previous sections are shown. For each point, the
singlet quartic coupling is determined from
$\lambda_S=\lambda_S^{\rm base}+a$, with $a$ sampled within the theoretically
allowed interval specified in Sec.~\ref{subsec:ewpt_numerics}.
}
    \label{fig:ResultsEWPTC1}
\end{figure}

In this subsection, the classification of a point as strong refers to the critical-temperature order parameter and does not by itself guarantee transition completion. Figure~\ref{fig:ResultsEWPTC1} shows the electroweak phase-transition pattern obtained for Scenario~I, the thermal two-component regime with $M_\psi<M_S<2M_\psi$. The blue points satisfy the theoretical and dark-matter constraints discussed in the previous sections, whereas green points also satisfy the strong-transition criterion.

For Scenario~I, a subset of the dark-matter viable parameter space satisfies the strong-transition criterion. These points are mainly associated with positive and sizeable values of the Higgs-portal coupling $\lambda_{HS}$, but their distribution also depends significantly on the dark sector parameters. This is consistent with the role of the portal in coupling the Higgs and singlet directions in the finite-temperature potential: a sufficiently active portal, together with an appropriate singlet mass scale, can support a barrier between the high-temperature phase and the electroweak phase. As the scalar mass increases, larger values of $\lambda_{HS}$ are typically required for the singlet sector to remain relevant during the thermal evolution.

The same trend appears in the $(M_\psi,\lambda_{HS})$ plane. Points satisfying
$\zeta_c\gtrsim1$ are concentrated at positive $\lambda_{HS}$ and relatively
light dark-sector masses. As $M_\psi$ increases, the strong-transition region
moves towards larger portal couplings and becomes more restricted. In the
Yukawa plane, the strong-transition points occupy a narrower region, showing
that the dark-sector interactions remain constrained once relic-density,
direct-detection, and phase-transition requirements are imposed simultaneously.

Scenario~II behaves differently. Although it is also a thermal two-component
regime, the lighter scalar component is strongly constrained by direct
detection. The surviving points are driven towards the Higgs-resonance region
and towards small Higgs-portal couplings. In this region, the singlet direction
has too little impact on the finite-temperature scalar potential to generate a
sizeable barrier between the relevant phases. Within the parameter space
considered, the points that survive the dark-matter constraints therefore do
not satisfy the strong-transition criterion along the Higgs direction.

A similar obstruction occurs in Scenario~III, the mixed two-component
WIMP--FIMP regime with a stable scalar component. Since the scalar survives
until today, it remains subject to direct-detection limits. As a result, the
dark-matter viable points tend to favour small Higgs-portal couplings or to
cluster around the Higgs-resonance region, reducing the portion of parameter
space in which the portal can efficiently contribute to the thermal barrier.
However, the absence of a strong transition in this scenario is not determined
by $\lambda_{HS}$ alone, but by its correlation with the other dark sector parameters. Within the dark-matter viable region explored
in our scan, these combined constraints do not lead to a strong first-order
electroweak phase transition according to $\zeta_c\gtrsim1$.

\begin{figure}[!htb]
    \centering

    \begin{minipage}{0.47\textwidth}
        \centering
        \includegraphics[width=\textwidth]{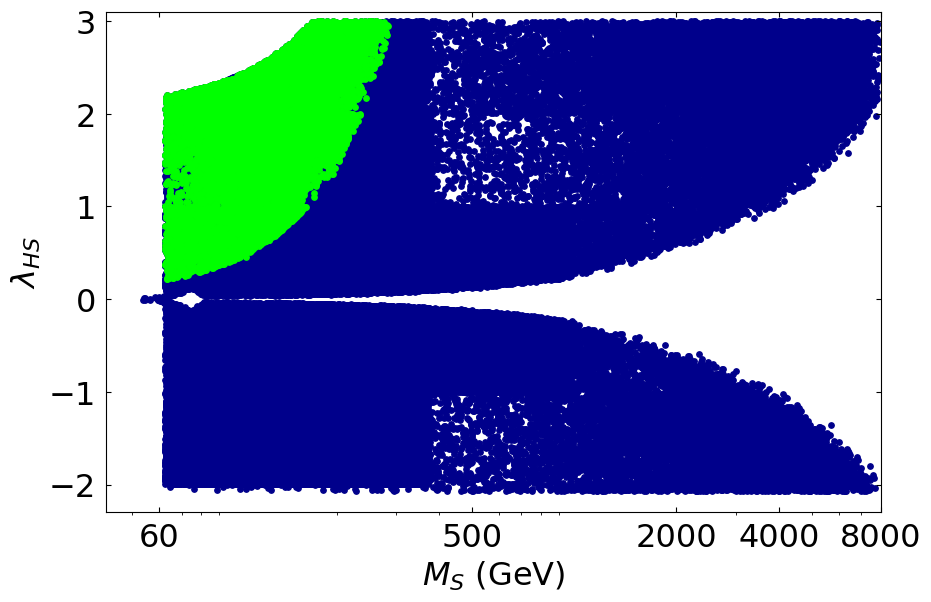}
    \end{minipage}\hfill
    \begin{minipage}{0.47\textwidth}
        \centering
        \includegraphics[width=\textwidth]{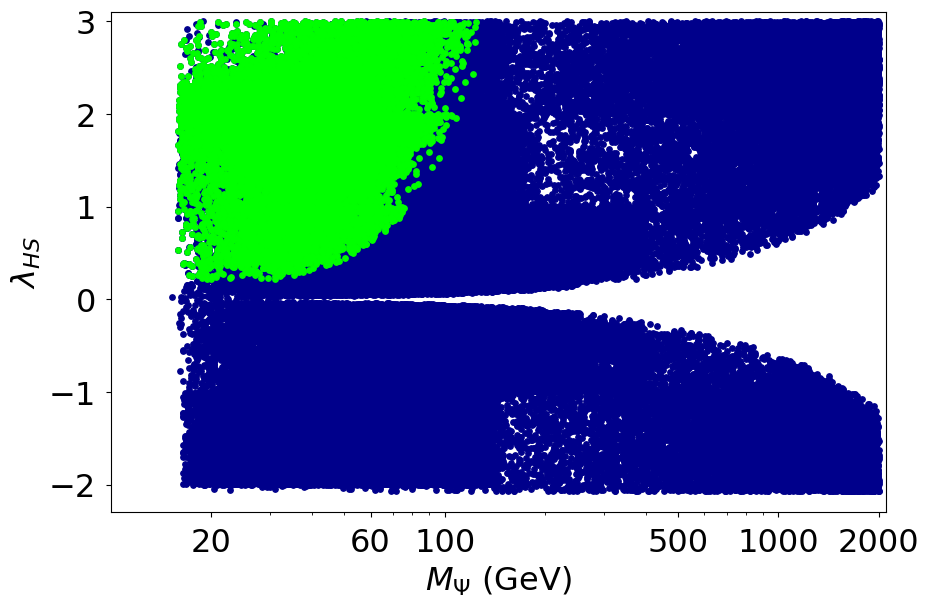}
    \end{minipage}

\caption{
Dependence of the electroweak phase-transition strength on the
parameters of Scenario~IV, defined by $M_S>2M_\psi$. In this effectively
one-component WIMP--FIMP regime, the scalar is thermally produced but unstable,
leaving the fermion as the surviving dark-matter component. The colour coding
separates points with $\zeta_c=v_c/T_c\gtrsim 1$ (green) from those with a weaker
transition (blue). The displayed sample has already been restricted by all
theoretical and dark-matter constraints. The point-dependent value of
$\lambda_S$ follows from $\lambda_S=\lambda_S^{\rm base}+a$, using the sampling
prescription given in Sec.~\ref{subsec:ewpt_numerics}.
}
    \label{fig:ResultsEWPTC4}
\end{figure}

Scenario~IV avoids this restriction. In this regime, the scalar is thermally
produced but unstable, with $M_S>2M_\psi$, and decays into the fermionic FIMP
component. The present-day dark matter is therefore fermionic, while the scalar
does not contribute to the halo abundance. Direct-detection constraints do not
force the Higgs-portal coupling to be small in the same way as in scenarios
with a stable scalar component. At the same time, the scalar is present in the
early thermal bath and enters the finite-temperature effective potential. This
allows a larger region in which the singlet direction can support a strong
first-order electroweak transition, as shown in
Fig.~\ref{fig:ResultsEWPTC4}.

\FloatBarrier

The correlation between the portal coupling and the scalar mass is again
visible in Scenario~IV. Larger scalar masses typically require larger
$\lambda_{HS}$ for the singlet direction to remain active during the transition.
Compared with Scenario~I, however, Scenario~IV allows a wider range of masses
and portal couplings because the scalar is not a stable dark-matter component.
Consequently, the region satisfying the strong-transition criterion is larger,
especially for positive and sizeable values of $\lambda_{HS}$.

\section{Gravitational Waves}
\label{sec:GW}

A first-order electroweak phase transition can source a stochastic
gravitational-wave background if bubbles of the broken phase nucleate and
expand in the surrounding plasma~\cite{Mazumdar2019,Caprini:2019egz}. The
critical temperature $T_c$ introduced in Sec.~\ref{sec:EWPT} marks the
degeneracy between the relevant phases, but the transition proceeds only after
bubble nucleation becomes efficient. For the benchmark spectra considered
below, we use the nucleation temperature $T_n$ as the reference temperature.

\subsection{Bubble nucleation and transition parameters}
\label{subsec:gw_nucleation}

Below $T_c$, the false vacuum can decay through thermal tunnelling. The
nucleation rate per unit volume is written as
\cite{Coleman:1977py,Callan:1977pt,Linde:1980tt,Quiros:1999jp}
\begin{equation}
\Gamma(T)
\simeq
T^4
\left(\frac{S_3(T)}{2\pi T}\right)^{3/2}
\exp\!\left[-\frac{S_3(T)}{T}\right],
\label{eq:nucleation_rate}
\end{equation}
where $S_3(T)$ is the three-dimensional Euclidean action of the
$O(3)$-symmetric bounce. In the present two-field system,
\begin{equation}
\boldsymbol{\Phi}(r)=\bigl(\phi(r),\sigma(r)\bigr),
\end{equation}
the action is
\begin{equation}
S_3(T)
=
4\pi
\int_0^\infty dr\, r^2
\left[
\frac{1}{2}
\frac{d\Phi_i}{dr}
\frac{d\Phi_i}{dr}
+
V_{\rm eff}\!\left(\boldsymbol{\Phi},T\right)
-
V_{\rm eff}\!\left(\boldsymbol{\Phi}^{\rm false},T\right)
\right],
\label{eq:S3_action}
\end{equation}
where repeated field-space indices are summed over. The subtraction of the
false-vacuum energy makes the action finite. The bounce profile satisfies
\begin{equation}
\frac{d^2\Phi_i}{dr^2}
+
\frac{2}{r}\frac{d\Phi_i}{dr}
=
\frac{\partial V_{\rm eff}}{\partial \Phi_i},
\label{eq:bounce_eom}
\end{equation}
with boundary conditions
\begin{equation}
\left.\frac{d\Phi_i}{dr}\right|_{r=0}=0,
\qquad
\lim_{r\to\infty}\Phi_i(r)=\Phi_i^{\rm false}.
\label{eq:bounce_bc}
\end{equation}

The nucleation temperature is estimated by requiring approximately one
critical bubble per Hubble volume. During radiation domination this condition
is commonly expressed as~\cite{Linde:1981zj,Quiros:1999jp,Mazumdar2019}
\begin{equation}
\frac{S_3(T_n)}{T_n}\simeq 140,
\label{eq:nucleation_condition}
\end{equation}
which we use as the practical criterion for identifying $T_n$ in the numerical
analysis. This approximation is adequate for the benchmark-level estimates
presented here; a full percolation analysis would require tracking the
false-vacuum fraction and the bubble-growth history.

The gravitational-wave spectrum is mainly controlled by the strength of the
transition and by its inverse time duration
~\cite{Caprini:2019egz,Schmitz:2020sjk}. We define the strength parameter as
\begin{equation}
\alpha
=
\frac{\Delta\theta(T_n)}
{\rho_{\rm rad}(T_n)},
\qquad
\rho_{\rm rad}(T_n)
=
\frac{\pi^2}{30}g_* T_n^4,
\label{eq:alpha_definition}
\end{equation}
where $g_*$ is the effective number of relativistic degrees of freedom at
$T_n$. The quantity $\Delta\theta$ is the difference in the trace-anomaly
contribution between the false and true phases,
\begin{equation}
\Delta\theta(T_n)
=
\left[
V_{\rm eff}(\boldsymbol{\Phi},T_n)
-
\frac{T_n}{4}
\left.
\frac{\partial V_{\rm eff}(\boldsymbol{\Phi},T)}
{\partial T}
\right|_{T_n}
\right]_{\boldsymbol{\Phi}^{\rm true}}^{\boldsymbol{\Phi}^{\rm false}},
\label{eq:trace_anomaly_definition}
\end{equation}
with the convention
\begin{equation}
[X]_{\boldsymbol{\Phi}^{\rm true}}^{\boldsymbol{\Phi}^{\rm false}}
\equiv
X(\boldsymbol{\Phi}^{\rm false})
-
X(\boldsymbol{\Phi}^{\rm true}).
\end{equation}
The temperature derivative in Eq.~\eqref{eq:trace_anomaly_definition} is taken
at fixed background fields.

The inverse time scale of the transition is quantified by
\begin{equation}
\frac{\beta}{H_n}
=
T_n
\left.
\frac{d}{dT}
\left(\frac{S_3}{T}\right)
\right|_{T_n},
\label{eq:beta_definition}
\end{equation}
with $H_n\equiv H(T_n)$. Larger values of $\alpha$ and smaller values of
$\beta/H_n$ generally enhance the gravitational-wave signal.

\subsection{Gravitational-wave spectrum}
\label{subsec:gw_spectrum}

The gravitational-wave background generated by a first-order phase transition
can receive contributions from scalar-field gradients, sound waves in the
plasma, and magnetohydrodynamic turbulence. In the following, we assume a
non-runaway regime in which most of the released energy is transferred to the
plasma rather than stored in the scalar-field configuration. We therefore
include the sound-wave and turbulence contributions,
\begin{equation}
\Omega_{\rm GW}h^2(f)
=
\Omega_{\rm sw}h^2(f)
+
\Omega_{\rm turb}h^2(f),
\label{eq:omega_gw_total}
\end{equation}
while neglecting the scalar-field bubble-collision contribution. The explicit
spectral functions used in the numerical analysis are collected in
Appendix~\ref{app:gw_spectral_functions}.

The sound-wave contribution is controlled by the efficiency factor
$\kappa_{\rm sw}$, which gives the fraction of the released energy converted
into bulk motion of the plasma. The turbulent component is parametrised by
$\kappa_{\rm turb}$ and mainly affects the tails of the spectrum \cite{berganholi2025cosmological}.
Following the prescription implemented in \texttt{PhaseTracer2}, we take
\begin{equation}
\kappa_{\rm turb}=0.1\,\kappa_{\rm sw}.
\label{eq:kappa_turb_choice}
\end{equation}
This choice reflects the standard assumption that turbulence receives only a
subdominant fraction of the bulk kinetic energy. The resulting signal is mainly
determined by $\alpha$, $\beta/H_n$, and $T_n$, together with the bubble-wall velocity $v_w$.

A first-principles determination of $v_w$ requires solving the
out-of-equilibrium plasma dynamics across the bubble wall, including the
friction exerted by particles in the thermal bath
~\cite{moore1995fast,de2022bubble}. This lies beyond the scope of the present
work. We therefore adopt
\begin{equation}
v_w=1,
\label{eq:wall_velocity_choice}
\end{equation}
as a relativistic-wall benchmark. This choice should not be interpreted as a
model-specific determination of the wall velocity; rather, it fixes the
kinematic input used to estimate the gravitational-wave spectra. Different wall
velocities would shift the peak frequency and amplitude of the predicted
signal.

The omission of the scalar-field collision contribution is consistent with the
same non-runaway assumption. Establishing this regime from first principles
would require a dedicated calculation of the wall dynamics and friction. Under
the benchmark assumptions above, Eq.~\eqref{eq:omega_gw_total} gives the
gravitational-wave signal associated with the first-order transitions selected
from the finite-temperature analysis.

In the benchmark analysis below, we use $T_n$ as the input temperature for the
spectral estimates. This provides a practical reference temperature and is
commonly adopted in benchmark-level scans. However, in supercooled transitions,
using $T_n$ instead of the percolation temperature $T_p$ can introduce a
systematic uncertainty, because the gravitational-wave signal depends on the
temperature at which a sufficient fraction of space has converted to the true
vacuum~\cite{athron2023supercool}. A complete treatment would require tracking
the cosmological history of the transition, including bubble nucleation,
bubble growth, and the false-vacuum fraction
~\cite{athron2023supercool,gg2020phase}. Studies of singlet extensions indicate
that a sizeable transition strength does not necessarily imply a large
hierarchy between $T_n$ and $T_p$, supporting the use of $T_n$ as a benchmark
reference in singlet-like scenarios~\cite{alves2020di}. We therefore interpret
the spectra below as indicative estimates rather than precision predictions,
and leave a dedicated percolation analysis for future work.

\subsection{Benchmark points and detector prospects}
\label{subsec:gw_benchmarks}

We now evaluate the gravitational-wave spectra for representative benchmark
points selected from the regions that satisfy the strong-transition criterion
and admit bubble nucleation. After imposing the theoretical and dark-matter
constraints, this occurs only in Scenarios~I and IV. Scenarios~II and III do
not contain points satisfying the strong-transition criterion within the viable
regions explored in our scan, and are therefore not included in the
gravitational-wave benchmark analysis.

For each benchmark point, the model parameters
\[
\{ M_S,\; M_\psi,\; y_s,\; y_p, \; \lambda_{HS},\; \lambda_{S} \}
\]
fix the finite-temperature effective potential and determine the transition
parameters $T_c$, $T_n$, $\alpha$, $\beta/H_n$, and $\zeta_c$. These quantities
set the characteristic frequency and amplitude of the gravitational-wave
spectrum under the assumptions specified in Sec.~\ref{subsec:gw_spectrum}. All
benchmark points listed below satisfy the theoretical and dark-matter
constraints, pass the strong-transition criterion $\zeta_c\gtrsim1$, and admit
a nucleation temperature according to Eq.~\eqref{eq:nucleation_condition}. The
values of $\lambda_S$ shown in the tables are those used in the corresponding
finite-temperature computation, following the vacuum-preserving prescription
defined in Sec.~\ref{subsec:vacuum_structure}.

The benchmark points for Scenario~I are shown in
Table~\ref{tab:bps_scenario_I}. In this regime, both dark-sector particles are
stable and thermally produced. The selected points illustrate the subset of the
thermal two-component parameter space in which the Higgs-portal coupling
remains large enough to support a strong electroweak transition while remaining
compatible with the dark-matter constraints.

\begin{table}[H]
\centering
\small
\setlength{\tabcolsep}{3.6pt}
\renewcommand{\arraystretch}{1.25}
\caption{
Selected benchmark points for Scenario~I, corresponding to the
thermal two-component regime $M_\psi<M_S<2M_\psi$. Model input parameters
are collected in the first block, followed by the thermal quantities obtained
from the finite-temperature effective potential. All dimensionful quantities
are expressed in GeV. Every benchmark satisfies the theoretical and
dark-matter constraints, fulfils the strong-transition condition
$\zeta_c\gtrsim 1$, and meets the bubble-nucleation criterion adopted in the
numerical analysis.
}
\begin{tabular}{@{}ccccccc@{\hspace{0.45cm}}ccccc@{}}
\toprule
&
\multicolumn{6}{c}{\textbf{Model inputs}}
&
\multicolumn{5}{c}{\textbf{Thermal outputs}} \\
\cmidrule(lr){2-7}
\cmidrule(lr){8-12}
\textbf{BP} &
\textbf{$M_\psi$} &
\textbf{$M_S$} &
\textbf{$y_s$} &
\textbf{$y_p$} &
\textbf{$\lambda_{HS}$} &
\textbf{$\lambda_S$} &
\textbf{$T_c$} &
\textbf{$T_n$} &
\textbf{$\alpha$} &
\textbf{$\beta/H_n$} &
\textbf{$\zeta_c$} \\
\midrule
1 & 137.600 & 170.950 & 0.6402 & 1.1871 & 2.1843 & 5.0184 & 100.18 & 64.11 & 0.078 & 211.21 & 2.10 \\
2 & 131.170 & 154.570 & 0.5230 & 0.8675 & 1.9895 & 4.6535 & 97.67 & 68.66 & 0.056 & 433.15 & 2.20 \\
3 & 162.420 & 213.530 & -0.1423 & -0.8900 & 2.2724 & 2.8136 & 112.73 & 81.40 & 0.036 & 275.23 & 1.64 \\
4 & 139.740 & 173.410 & 0.5385 & 0.9815 & 2.4098 & 6.9549 & 103.98 & 85.82 & 0.022 & 973.04 & 1.95 \\
5 & 140.840 & 158.860 & -0.3514 & -0.2852 & 1.1445 & 0.7665 & 124.03 & 111.79 & 0.008 & 1683.52 & 1.27 \\
6 & 151.160 & 190.800 & 0.1194 & 0.7379 & 2.1553 & 5.2892 & 120.03 & 112.21 & 0.007 & 3110.31 & 1.38 \\
7 & 154.310 & 193.670 & 0.3135 & -0.4849 & 2.2314 & 6.2863 & 122.24 & 116.55 & 0.005 & 4646.94 & 1.29 \\
\bottomrule
\end{tabular}
\label{tab:bps_scenario_I}
\end{table}

\begin{figure}[!htbp]
    \centering
    \begin{minipage}{0.55\textwidth}
        \centering
         \includegraphics[width=\textwidth]{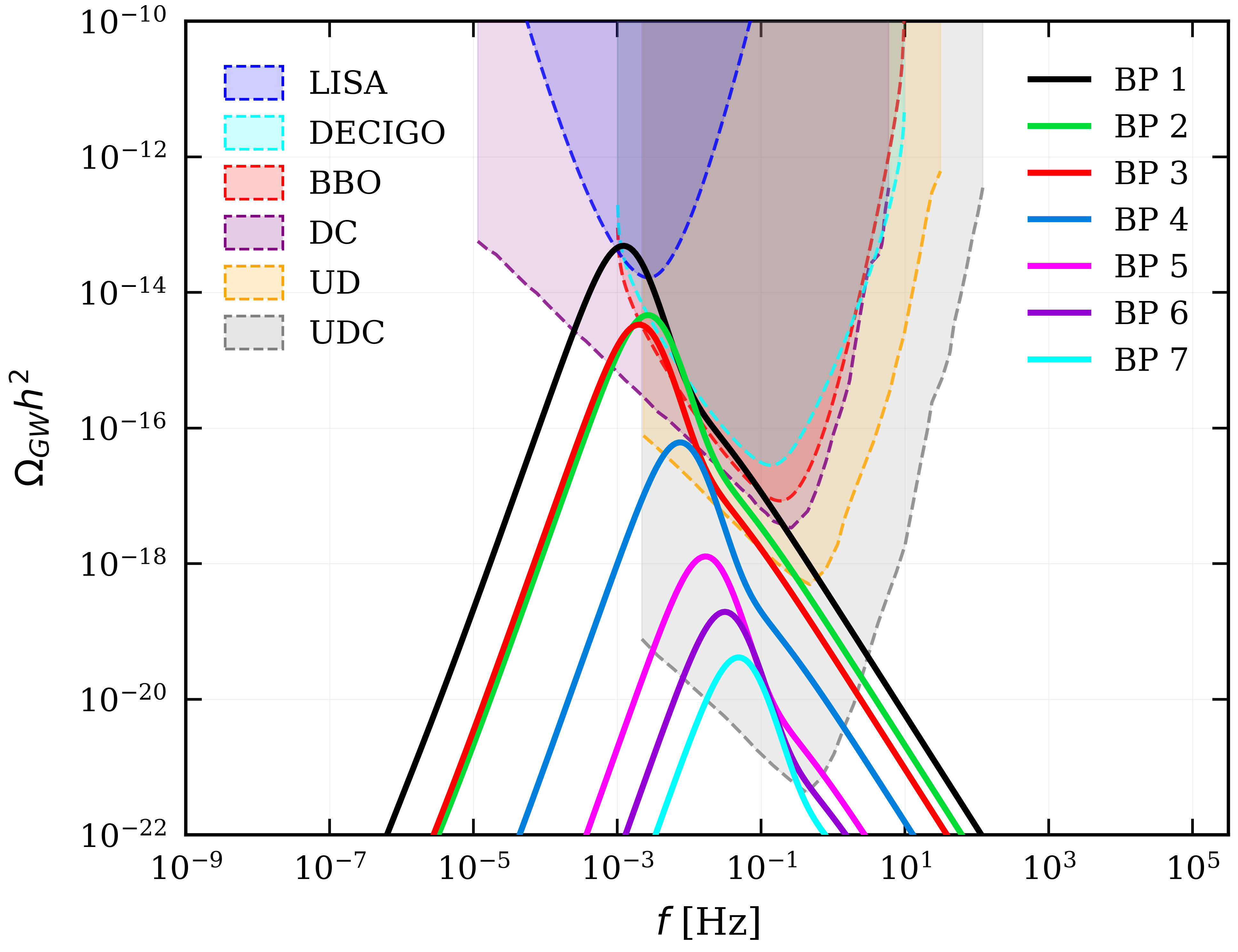}
    \end{minipage}

   \caption{
Predicted stochastic gravitational-wave spectra for the benchmark points in
Scenario~I, with $M_\psi<M_S<2M_\psi$. The spectra include the sound-wave and
turbulence contributions computed with the benchmark choice $v_w=1$, and are
compared with the projected sensitivity curves shown in the figure, including
LISA~\cite{LISA:2017pwj}, BBO~\cite{Yagi:2011va}, and the DECIGO-related
configurations DECIGO, UD, DC, and
UDC~\cite{Kawamura:2021decigo,Nakayama:2009ce}.
}
\label{fig:gw_scenario_I}
\end{figure}

The corresponding gravitational-wave spectra are shown in
Fig.~\ref{fig:gw_scenario_I}. These curves illustrate how the transition
parameters listed in Table~\ref{tab:bps_scenario_I} translate into the
present-day energy density $\Omega_{\rm GW}h^2$ as a function of frequency. In
this scenario, the largest signals are not determined by $\alpha$ alone, but by
the combined effect of the transition strength, the inverse duration
$\beta/H_n$, and the nucleation temperature. For instance, BP1 produces the largest peak among the benchmark points, whereas the spectrum for BP7 is strongly suppressed. Moreover, BP2 and BP3 provide a more instructive comparison: BP2 has a larger transition strength, $\alpha=0.056$ compared with $0.036$ for BP3, but also a larger $\beta/H_n$, $433.15$ compared with $275.23$, resulting in rather similar spectra.

The benchmark points for Scenario~IV are shown in
Table~\ref{tab:bps_scenario_IV}. In this regime, $M_S>2M_\psi$, so the scalar
is thermally produced but unstable, and the present-day dark matter is purely
fermionic. The small Yukawa couplings keep the fermion out of equilibrium,
while the Higgs portal controls the scalar thermal history and its effect on
the finite-temperature potential. Scenario~IV is therefore relevant for
gravitational-wave phenomenology because the scalar can affect the electroweak
thermal dynamics without surviving today as a directly constrained scalar
dark-matter component.

\begin{table}[!htbp]
\centering
\small
\setlength{\tabcolsep}{3.6pt}
\renewcommand{\arraystretch}{1.25}
\caption{
Scenario~IV is illustrated by benchmark configurations in the
effectively one-component WIMP--FIMP regime, defined by $M_S>2M_\psi$.
The upper part of the table specifies the model parameters, whereas the lower
part reports the corresponding finite-temperature results. Masses and
temperatures are expressed in GeV, with the Yukawa couplings quoted in units
of $10^{-12}$. The displayed points obey all theoretical and dark-matter
constraints and lead to successful bubble nucleation, in addition to
satisfying the strong-transition requirement $\zeta_c\gtrsim 1$.
}
\begin{tabular}{@{}ccccccc@{\hspace{0.45cm}}ccccc@{}}
\toprule
&
\multicolumn{6}{c}{\textbf{Model inputs}}
&
\multicolumn{5}{c}{\textbf{Thermal outputs}} \\
\cmidrule(lr){2-7}
\cmidrule(lr){8-12}
\textbf{BP} &
\textbf{$M_\psi$} &
\textbf{$M_S$} &
\textbf{$y_s\,[10^{-12}]$} &
\textbf{$y_p\,[10^{-12}]$} &
\textbf{$\lambda_{HS}$} &
\textbf{$\lambda_S$} &
\textbf{$T_c$} &
\textbf{$T_n$} &
\textbf{$\alpha$} &
\textbf{$\beta/H_n$} &
\textbf{$\zeta_c$} \\
\midrule
1 & 33.187 & 83.481 & -1.8063 & -1.8063 & 0.4570 & 0.2000 & 100.46 & 48.14 & 0.303 & 99.98 & 2.13 \\
2 & 31.936 & 67.345 & -7.7092 & 1.1685 & 0.4617 & 0.2883 & 85.99 & 37.70 & 0.560 & 265.78 & 2.69 \\
3 & 22.327 & 65.808 & -2.2786 & 1.3718 & 0.4451 & 0.2766 & 87.23 & 47.88 & 0.219 & 382.09 & 2.64 \\
4 & 29.561 & 221.350 & -2.3347 & 2.2828 & 2.7149 & 5.2270 & 109.61 & 87.39 & 0.024 & 390.37 & 1.72 \\
5 & 32.732 & 90.648 & 1.8355 & 1.8355 & 0.4386 & 0.1540 & 113.24 & 86.95 & 0.030 & 673.59 & 1.69 \\
6 & 38.791 & 84.860 & 4.2977 & -1.9400 & 0.3945 & 0.1474 & 115.35 & 96.47 & 0.018 & 1138.49 & 1.62 \\
7 & 107.510 & 242.440 & -2.7545 & -2.1830 & 2.6575 & 3.7648 & 123.40 & 113.98 & 0.007 & 2062.03 & 1.22 \\
\bottomrule
\end{tabular}
\label{tab:bps_scenario_IV}
\end{table}

\begin{figure}[!htbp]
    \centering
    \begin{minipage}{0.55\textwidth}
        \centering
        \includegraphics[width=\textwidth]{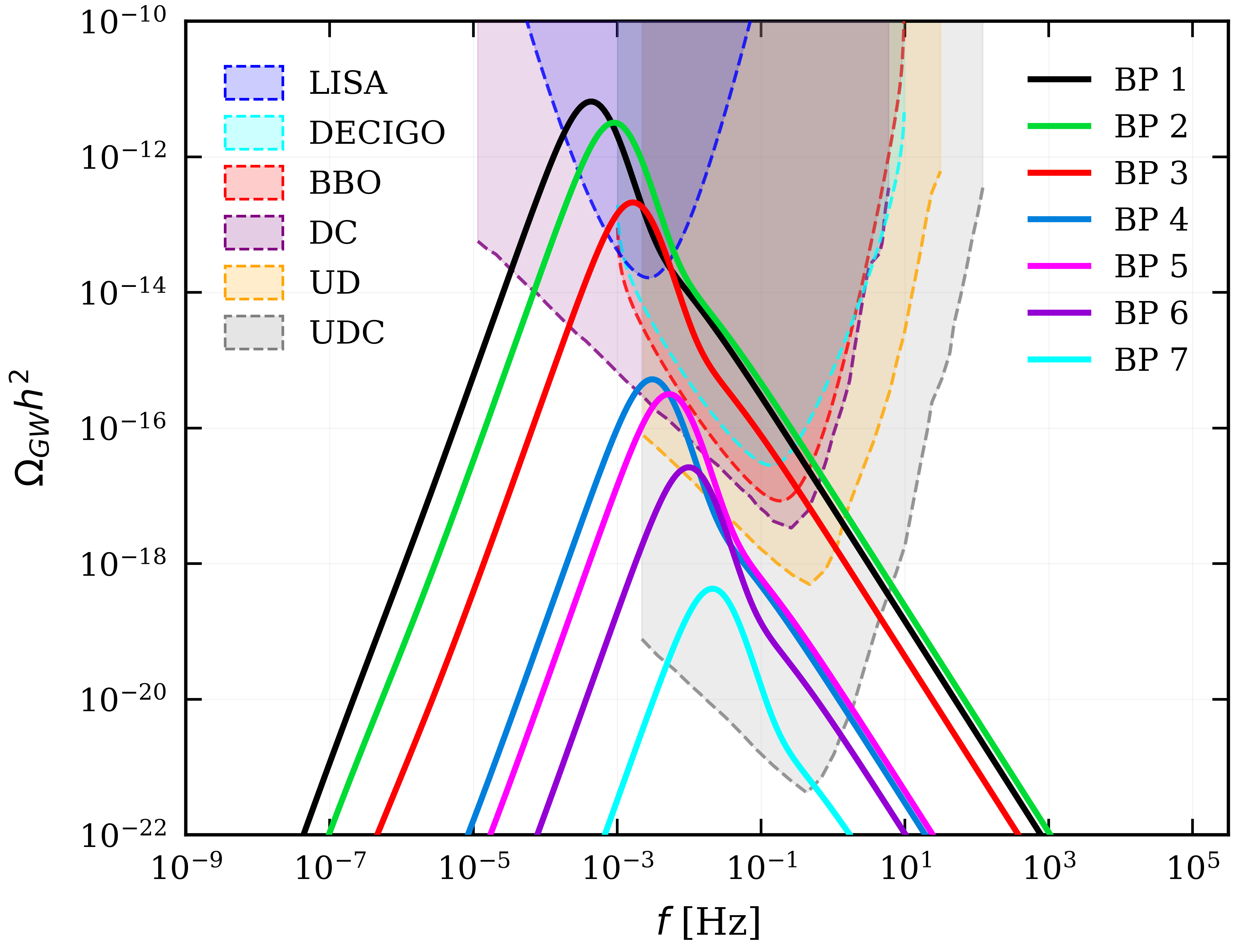}
    \end{minipage}

  \caption{
Predicted stochastic gravitational-wave spectra for the benchmark points in
Scenario~IV, with $M_S>2M_\psi$. The spectra include the sound-wave and
turbulence contributions computed with the benchmark choice $v_w=1$, and are
compared with the projected sensitivity curves shown in the figure, including
LISA~\cite{LISA:2017pwj}, BBO~\cite{Yagi:2011va}, and the DECIGO-related
configurations DECIGO, UD, DC, and
UDC~\cite{Kawamura:2021decigo,Nakayama:2009ce}.
}
\label{fig:gw_scenario_IV}
\end{figure}

The separation between the critical and nucleation temperatures is
particularly pronounced for BP1--BP3. Defining the degree of supercooling as
$\delta_{\rm sc}\equiv 1-T_n/T_c$, these points give
$\delta_{\rm sc}\simeq 0.52$, $0.56$, and $0.45$, respectively. These benchmarks also yield the most prominent gravitational-wave spectra in Scenario IV, as shown in Fig.~\ref{fig:gw_scenario_IV}. The corresponding gravitational-wave spectra should therefore
be interpreted as benchmark estimates subject to the percolation-temperature
uncertainty discussed in Sec.~\ref{subsec:gw_spectrum}.

Compared with Scenario~I, Scenario~IV is less
restricted by direct detection because the scalar does not survive as a
present-day dark-matter component. Sizeable Higgs-portal couplings can
therefore remain compatible with dark-matter constraints, allowing the scalar
sector to support a strong first-order transition while the final relic
abundance is fermionic.

Within this benchmark set, the largest peak amplitudes are obtained
for points combining relatively large $\alpha$ with smaller $\beta/H_n$. This
behaviour follows from the dependence of the sound-wave and turbulence
contributions on the efficiency-weighted transition strength and on the
transition duration. The comparison with the projected sensitivities also
depends on the peak frequency, which is controlled mainly by $T_n$ and
$\beta/H_n$ for fixed $v_w$. Accordingly, some benchmark spectra overlap the
projected sensitivity curves shown in Fig.~\ref{fig:gw_scenario_IV}, whereas
points with weaker transitions, larger $\beta/H_n$, or both generally remain
below them.

Taken together, the benchmark spectra show that viable dark matter, a strong
first-order electroweak transition, bubble nucleation, and gravitational-wave
signals within the reach of future interferometers can coexist in selected
regions of the model. This coexistence is not generic. It occurs when the Higgs
portal remains large enough for the singlet sector to affect the electroweak
thermal history while still satisfying the constraints imposed by present
dark-matter phenomenology. This correlation is the central link between the
dark-sector history, the electroweak phase transition, and the resulting
gravitational-wave signal.

\FloatBarrier

\section{Conclusions}
\label{sec:conclusions}

We have analysed a minimal $\mathbb{Z}_4$-symmetric fermion--scalar extension
of the SM in which dark-matter production, electroweak phase-transition
dynamics, and gravitational-wave signals are connected through a small set of
dark-sector parameters. Although the field content is economical, the
$\mathbb{Z}_4$ charge assignment allows qualitatively different cosmological
histories, including thermal two-component dark matter, mixed WIMP--FIMP
production, and an effectively fermionic relic abundance generated by scalar
decays.

The main result is that the simultaneous realisation of viable dark matter and
a strong electroweak phase transition is highly selective. After imposing the
zero-temperature consistency conditions and the dark-matter constraints, strong
transitions are found only in Scenarios~I and IV within the parameter space
considered. In Scenario~I, with $M_\psi<M_S<2M_\psi$, both dark-sector particles
are stable and thermally produced, but the scalar component can remain
sufficiently subdominant for sizeable Higgs-portal couplings to survive direct
detection. In Scenario~IV, with $M_S>2M_\psi$, the scalar is thermally produced
but unstable; the present-day relic abundance is fermionic, while the scalar
sector can still affect the electroweak thermal potential.

By contrast, Scenarios~II and III do not yield strong electroweak transitions after the dark-matter constraints are imposed. In both cases, the scalar is a stable component of the present-day dark matter and is therefore strongly constrained by Higgs-mediated direct detection. The surviving points tend to favour small Higgs-portal couplings or to lie near the Higgs-resonance region, thereby reducing the range of parameter space in which the singlet sector can efficiently contribute to the finite-temperature barrier. The difference between successful and unsuccessful regimes is therefore not determined by the Higgs portal alone, but by its interplay with the scalar mass, the mass hierarchy, and the remaining singlet-sector parameters under the present dark-matter bounds.

The viable strong transitions are typically two-step transitions,
\[
(0,0)\to(0,w)\to(v',0),
\]
where an intermediate singlet-like phase separates the high-temperature phase
from the electroweak vacuum. For representative benchmark points that also
admit bubble nucleation, we computed the corresponding stochastic
gravitational-wave spectra from sound waves and turbulence in the plasma. Under
the benchmark assumptions adopted for the wall velocity, plasma dynamics, and
nucleation temperature, some spectra enter the projected sensitivity range of
future space-based interferometers.

The resulting picture is therefore not a generic prediction of observable gravitational waves, but a correlated selection of parameter regions. Present dark-matter constraints restrict the combinations of parameters for which the scalar and fermion sectors can remain relevant to the electroweak thermal history. Future gravitational-wave searches can therefore probe regions of the model in which the same dark-sector interactions that shape the relic history also contribute efficiently to the finite-temperature dynamics.

A natural next step is to refine the quantitative predictions through a
first-principles determination of the bubble-wall velocity, a dedicated
percolation analysis, and a two-field sphaleron calculation. The qualitative
picture found here remains unchanged: in the $\mathbb{Z}_4$ fermion--scalar
framework, dark-matter viability, a strong two-step electroweak transition, and
gravitational-wave signals within the reach of future interferometers can
coexist, but only in selected regions of parameter space, predominantly those
represented by Scenarios~I and IV.

\appendix

\section{Tree-level vacuum structure and perturbative unitarity}
\label{app:zero_temperature_details}

\subsection{Perturbative unitarity in the scalar sector}
\label{app:unitarity_details}

We briefly derive the perturbative-unitarity conditions used in
Sec.~\ref{sec:constraints}. In the high-energy limit, scalar $2\to2$
amplitudes are controlled by the quartic interactions, since contributions
from dimensionful parameters are suppressed by powers of the centre-of-mass
energy. The corresponding $s$-wave amplitudes can therefore be organised into a
constant scattering matrix.

In an orthonormal basis of scalar two-body states, the scattering matrix
contains uncoupled channels and a nontrivial $2\times2$ block,
\begin{equation}
\mathcal{Q}=
\begin{pmatrix}
6\lambda_H & \lambda_{HS}\\
\lambda_{HS} & 3\lambda_S
\end{pmatrix}.
\label{eq:Q_unitarity_app}
\end{equation}
The eigenvalues of this block are
\begin{equation}
\Lambda_\pm
=
\frac{1}{2}\left(
6\lambda_H+3\lambda_S
\pm
\sqrt{36\lambda_H^2+9\lambda_S^2+4\lambda_{HS}^2-36\lambda_H\lambda_S}
\right).
\label{eq:Lambda_pm_app}
\end{equation}
The corresponding zeroth partial-wave amplitudes are
$a_0=\Lambda_\pm/(16\pi)$. The uncoupled channels give additional eigenvalues
proportional to $2\lambda_H$ and $\lambda_{HS}$, leading to the separate bounds
on $\lambda_H$ and $\lambda_{HS}$ quoted in the main text.

Imposing $|a_0|\leq 1/2$ on all eigenchannels gives
\begin{equation}
|\lambda_H|\leq 4\pi,
\qquad
|\lambda_{HS}|\leq 8\pi,
\qquad
|\Lambda_\pm|\leq 8\pi.
\label{eq:unitarity_bounds_app}
\end{equation}
These are the perturbative-unitarity conditions used in
Sec.~\ref{sec:constraints}. They provide the upper theoretical restriction on
the scalar quartics, and in particular on how large the singlet self-coupling
may be chosen once the vacuum-depth lower bound has been imposed.

\subsection{Zero-temperature stationary points and competing vacua}
\label{app:vacuum_structure_details}

We now summarise the tree-level vacuum analysis underlying
Sec.~\ref{subsec:vacuum_structure}. Along the neutral field directions, the
scalar potential is
\begin{equation}
V_0(h,s)
=
-\frac{1}{2}\mu_H^2 h^2
+\frac{1}{4}\lambda_H h^4
-\frac{1}{2}\mu_S^2 s^2
+\frac{1}{4}\lambda_S s^4
+\frac{1}{4}\lambda_{HS} h^2 s^2.
\label{eq:V0_hs_app}
\end{equation}
The stationarity conditions are
\begin{align}
\frac{\partial V_0}{\partial h}
&=
h\left(-\mu_H^2+\lambda_H h^2+\frac{1}{2}\lambda_{HS}s^2\right)=0,
\label{eq:dVdh_app}
\\
\frac{\partial V_0}{\partial s}
&=
s\left(-\mu_S^2+\lambda_S s^2+\frac{1}{2}\lambda_{HS}h^2\right)=0.
\label{eq:dVds_app}
\end{align}
The possible stationary points are the origin, the electroweak extremum
$(v,0)$ with $v^2=\mu_H^2/\lambda_H$, the pure singlet extremum $(0,v_s)$ with
$v_s^2=\mu_S^2/\lambda_S$ when $\mu_S^2>0$, and a mixed configuration
$(v_h,v_s)$ with both field values nonzero.

For the mixed configuration, Eqs.~\eqref{eq:dVdh_app} and
\eqref{eq:dVds_app} imply
\begin{equation}
-\mu_H^2+\lambda_H v_h^2+\frac{1}{2}\lambda_{HS}v_s^2=0,
\qquad
-\mu_S^2+\lambda_S v_s^2+\frac{1}{2}\lambda_{HS}v_h^2=0.
\label{eq:mixed_stat_app}
\end{equation}
Solving for the two field values gives
\begin{equation}
v_h^2=
\frac{4\lambda_S\mu_H^2-2\lambda_{HS}\mu_S^2}
{4\lambda_H\lambda_S-\lambda_{HS}^2},
\qquad
v_s^2=
\frac{4\lambda_H\mu_S^2-2\lambda_{HS}\mu_H^2}
{4\lambda_H\lambda_S-\lambda_{HS}^2}.
\label{eq:mixed_vevs_app}
\end{equation}

At the electroweak vacuum, the tree-level relations
\[
\mu_H^2=\lambda_H v^2,
\qquad
M_S^2=-\mu_S^2+\frac{1}{2}\lambda_{HS}v^2
\]
reparametrise the Lagrangian in terms of quantities defined at $(v,0)$. Once
$\lambda_H>0$ is imposed by boundedness from below, local stability of the
electroweak vacuum requires $M_S^2>0$. Using these relations in
Eq.~\eqref{eq:mixed_vevs_app}, the singlet component of the mixed stationary
point becomes
\begin{equation}
v_s^2=
-\frac{4\lambda_H M_S^2}{\mathcal{D}},
\qquad
\mathcal{D}\equiv 4\lambda_H\lambda_S-\lambda_{HS}^2.
\label{eq:vs2_reparam_app}
\end{equation}

This expression eliminates the mixed stationary point as an independent
competing vacuum. If $\mathcal{D}>0$, then $v_s^2<0$ and the mixed solution is
not physical. If $\mathcal{D}<0$, a physical mixed solution may exist, but it
cannot be a minimum. Indeed, the Hessian at $(v_h,v_s)$ is
\begin{equation}
\mathcal{H}(v_h,v_s)=
\begin{pmatrix}
2\lambda_H v_h^2 & \lambda_{HS}v_hv_s\\
\lambda_{HS}v_hv_s & 2\lambda_S v_s^2
\end{pmatrix},
\end{equation}
with determinant
\begin{equation}
\det \mathcal{H}(v_h,v_s)
=
v_h^2 v_s^2\left(4\lambda_H\lambda_S-\lambda_{HS}^2\right)
=
v_h^2 v_s^2\,\mathcal{D}.
\end{equation}
Thus, whenever a physical mixed solution exists in the branch
$\mathcal{D}<0$, the Hessian has negative determinant and the stationary point
is a saddle. The boundary $\mathcal{D}=0$ is degenerate and does not define an
additional competing minimum in the parameter regions retained in the scan.

Consequently, after imposing $M_S^2>0$, the only tree-level vacuum that can
compete with the electroweak one is the pure singlet extremum. If
$\mu_S^2\leq0$, this extremum is absent. If $\mu_S^2>0$, it is located at
\begin{equation}
v_s^2=\frac{\mu_S^2}{\lambda_S},
\end{equation}
with vacuum energy
\begin{equation}
V_0(0,v_s)=-\frac{\mu_S^4}{4\lambda_S}.
\end{equation}
The electroweak vacuum has
\begin{equation}
V_0(v,0)=-\frac{\mu_H^4}{4\lambda_H}.
\end{equation}
Requiring the electroweak vacuum to be deeper,
\begin{equation}
V_0(v,0)<V_0(0,v_s),
\end{equation}
gives
\begin{equation}
\lambda_S>
\lambda_H\frac{\mu_S^4}{\mu_H^4}.
\label{eq:lambdaS_bound_app_mu}
\end{equation}
Using the electroweak-vacuum parametrisation, this becomes
\begin{equation}
\lambda_S>\lambda_S^{\rm min},
\qquad
\lambda_S^{\rm min}
=
\frac{\left(M_S^2-\frac{1}{2}\lambda_{HS}v^2\right)^2}
{\lambda_H v^4},
\qquad
(\mu_S^2>0).
\label{eq:lambdaSmin_app}
\end{equation}

\section{Thermal self-energies and resummed masses}
\label{app:thermal_masses}

This appendix lists the thermal self-energies and resummed masses used in the
Parwani implementation of Sec.~\ref{sec:effective_potential}. For bosonic modes
receiving leading plasma corrections, we replace
\begin{equation}
m_i^2(\phi,\sigma)
\;\longrightarrow\;
\mathcal{M}_i^2(\phi,\sigma,T)
=
m_i^2(\phi,\sigma)+\Pi_i(T).
\end{equation}
Scalar modes and longitudinal gauge modes are diagonalised after the thermal
corrections are included. Transverse gauge modes and fermions are not
daisy-resummed.

\paragraph{Scalar sector.}

At leading order, the scalar thermal self-energies are field independent and
diagonal in the $(\phi,\sigma)$ basis:
\begin{equation}
\begin{split}
 \Pi_h(T)=&\Pi_G(T)=
\left(
\frac{3g^2}{16}
+\frac{g'^2}{16}
+\frac{y_t^2}{4}
+\frac{\lambda_H}{2}
+\frac{\lambda_{HS}}{24}
\right)T^2,
\\
\Pi_s(T)=&
\left(
\frac{\lambda_S}{4}
+\frac{\lambda_{HS}}{6}
+\frac{y_s^2+y_p^2}{6}
\right)T^2.
\label{eq:Pi_scalar_app}   
\end{split}
\end{equation}
No independent off-diagonal thermal self-energy is generated at this order, so
the mixing of the resummed scalar modes is inherited from the tree-level entry
$\mathcal{M}_{12}^2(\phi,\sigma)=\lambda_{HS}\phi\sigma$.

The thermally corrected scalar mass matrix is
\begin{equation}
\mathcal{M}_{\mathrm{scal},T}^2(\phi,\sigma,T)
=
\begin{pmatrix}
\mathcal{M}_{11}^2(\phi,\sigma)+\Pi_h(T) & \mathcal{M}_{12}^2(\phi,\sigma)\\[2mm]
\mathcal{M}_{12}^2(\phi,\sigma) & \mathcal{M}_{22}^2(\phi,\sigma)+\Pi_s(T)
\end{pmatrix},
\label{eq:MscalT_app}
\end{equation}
where $\mathcal{M}_{ij}^2(\phi,\sigma)$ are defined in
Eq.~\eqref{eq:scalar_mass_matrix}. Its eigenvalues are
\begin{equation}
\begin{split}
\mathcal{M}_{h_{1,2}}^2(\phi,\sigma,T)
=&
\frac{1}{2}
\left[
\mathcal{M}_{11}^2+\mathcal{M}_{22}^2+\Pi_h+\Pi_s \right .\\
&\left . \pm
\sqrt{
\left(\mathcal{M}_{11}^2-\mathcal{M}_{22}^2+\Pi_h-\Pi_s\right)^2
+
4\left(\mathcal{M}_{12}^2\right)^2
}
\right].
\label{eq:Mh12T_app}
\end{split}
\end{equation}
The Goldstone modes are resummed as
\begin{equation}
\mathcal{M}_G^2(\phi,\sigma,T)
=
m_G^2(\phi,\sigma)+\Pi_G(T),
\label{eq:MGT_app}
\end{equation}
with $m_G^2(\phi,\sigma)$ given in Eq.~\eqref{eq:goldstone_mass}.

The Yukawa contribution to $\Pi_s(T)$ follows from the field-dependent
singlet-fermion spectrum,
\begin{equation}
m_{\psi_1}^2(\sigma)+m_{\psi_2}^2(\sigma)
=
2M_\psi^2+2(y_s^2+y_p^2)\sigma^2,
\end{equation}
which gives the term $(y_s^2+y_p^2)T^2/6$ in
Eq.~\eqref{eq:Pi_scalar_app}.

\paragraph{Gauge sector.}

For the longitudinal charged gauge modes,
\begin{equation}
\Pi_{W_L}(T)=\frac{11}{6}g^2T^2,
\qquad
\mathcal{M}_{W_L}^2(\phi,T)
=
m_W^2(\phi)+\Pi_{W_L}(T)
=
\frac{g^2\phi^2}{4}+\frac{11}{6}g^2T^2.
\label{eq:WL_resummed_app}
\end{equation}

In the $(W_3,B)$ basis, the neutral longitudinal sector is described by
\begin{equation}
\mathcal{M}_{L}^2(\phi,T)=
\begin{pmatrix}
\dfrac{g^2\phi^2}{4}+\dfrac{11}{6}g^2T^2 & -\dfrac{gg'}{4}\phi^2 \\[3mm]
-\dfrac{gg'}{4}\phi^2 & \dfrac{g'^2\phi^2}{4}+\dfrac{11}{6}g'^2T^2
\end{pmatrix}.
\label{eq:neutral_gauge_matrix_app}
\end{equation}
Its eigenvalues are
\begin{align}
\mathcal{M}_{Z_L}^2(\phi,T)
=&
\frac{1}{2}
\left[
\frac{1}{4}(g^2+g'^2)\phi^2
+\frac{11}{6}(g^2+g'^2)T^2\right .\nonumber\\
&\left . +
\sqrt{
(g^2-g'^2)^2\left(\frac{\phi^2}{4}+\frac{11}{6}T^2\right)^2
+\frac{g^2g'^2}{4}\phi^4
}
\right],
\label{eq:MZLT_app}
\\[2mm]
\mathcal{M}_{\gamma_L}^2(\phi,T)
=&
\frac{1}{2}
\left[
\frac{1}{4}(g^2+g'^2)\phi^2
+\frac{11}{6}(g^2+g'^2)T^2\right .\nonumber\\
&\left . -
\sqrt{
(g^2-g'^2)^2\left(\frac{\phi^2}{4}+\frac{11}{6}T^2\right)^2
+\frac{g^2g'^2}{4}\phi^4
}
\right].
\label{eq:MgammaLT_app}
\end{align}

The transverse gauge modes keep their tree-level field-dependent masses. Thus,
in the gauge contribution to Eqs.~\eqref{eq:VCW_Parwani} and
\eqref{eq:VT_Parwani}, the charged modes are split into two longitudinal
degrees of freedom with mass $\mathcal{M}_{W_L}^2$ and four transverse degrees
of freedom with mass $m_W^2$. In the neutral sector, the longitudinal modes are
described by $\mathcal{M}_{Z_L}^2$ and $\mathcal{M}_{\gamma_L}^2$, while the
transverse modes retain their tree-level masses.

\paragraph{Fermionic sector.}

Fermions are not daisy-resummed in the Parwani prescription used here. Their
contribution to the effective potential is evaluated with the zero-temperature
field-dependent eigenvalues $m_{\psi_{1,2}}^2(\sigma)$ and $m_t^2(\phi)$
introduced in Sec.~\ref{sec:effective_potential}.

\paragraph{Degrees of freedom.}

In the numerical sums, the scalar modes $h_1$ and $h_2$ carry one degree of
freedom each, and the Goldstone sector carries three. The top quark carries
twelve fermionic degrees of freedom. The two singlet-fermion eigenstates carry
\begin{equation}
n_{\psi_1}=n_{\psi_2}=2,
\end{equation}
so that together they account for the four degrees of freedom of the original
Dirac fermion. Gauge degrees of freedom are split into longitudinal resummed
modes and transverse tree-level modes as described above.

\paragraph{Remark on complex contributions.}

Local imaginary parts can appear in regions where bosonic squared masses become
negative. As stated in Sec.~\ref{sec:effective_potential}, we use the real part
of the effective potential when tracing the relevant minima. This prescription
does not alter the resummed masses listed in this appendix.

 \section{Gravitational-wave spectral functions}
 \label{app:gw_spectral_functions}

This appendix collects the spectral functions used to compute the benchmark gravitational-wave spectra in Sec.~\ref{sec:GW}. We compute the benchmark spectra following the \texttt{PhaseTracer2} implementation, which uses the phenomenological parametrisations for sound-wave \cite{hindmarsh2015numerical, hindmarsh2020erratum,Caprini:2019egz} and magnetohydrodynamic-turbulence \cite{caprini2009stochastic,binetruy2012cosmological} contributions. The total spectrum is
\begin{equation}
\Omega_{\rm GW}h^2(f) = \Omega_{\rm sw}h^2(f) + \Omega_{\rm turb}h^2(f).
\end{equation}

All transition quantities are evaluated at the nucleation temperature $T_n$,
used as the benchmark reference temperature in Sec.~\ref{subsec:gw_spectrum}.
We denote

\begin{equation}
H_n\equiv H(T_n),
\qquad
R_n\equiv R(T_n) = (8\pi)^{1/3}\frac{v_w}{\beta},
\label{eq:app_Rn}
\end{equation}
where $v_w$ is the bubble-wall velocity, $\beta^{-1}$ characterises the duration of the transition and $R_n$ is the mean bubble separation.

\paragraph{Sound waves.}

The sound-wave contribution is evaluated as
\begin{multline}
\Omega_{\rm sw}h^2(f) = 4.0587\times 10^{-7} \left(\frac{100}{g_*}\right)^{1/3} \left(\frac{\kappa_{\rm sw}\alpha}{1+\alpha}\right)^2 \left(\frac{f}{f_{\rm sw}}\right)^3 \left[\frac{7}{4+3(f/f_{\rm sw})^2}\right]^{7/2}
\\[1mm]
\times
\min\!\left(\frac{H_n R_n}{\overline{U}_f},1\right)
H_n R_n .
\label{eq:app_Omega_sw}
\end{multline}
Here $\kappa_{\rm sw}$ is the efficiency factor for converting released vacuum energy into bulk motion of the plasma, and
\begin{equation}
\overline{U}_f =\left[\frac{3}{4}\frac{\kappa_{\rm sw}\alpha}{1+\alpha}\right]^{1/2}
\label{eq:app_Ubar}\end{equation}
is the root-mean-square fluid velocity. The characteristic sound-wave frequency is
\begin{equation}\frac{f_{\rm sw}}{1\,\mu{\rm Hz}}=26\left(\frac{T_n}{100\,{\rm GeV}}\right)\left(\frac{g_*}{100}\right)^{1/6}\frac{1}{H_n R_n}.
\label{eq:app_fsw}
\end{equation}

For the relativistic-wall benchmark used in the main text, we adopt the detonation efficiency fit \cite{espinosa2010energy}
\begin{equation}
\kappa_{\rm sw}
=
\frac{
(v_J-1)^3 v_J^{5/2} v_w^{-5/2}\kappa_C\kappa_D
}{
\left[(v_J-1)^3-(v_w-1)^3\right]v_J^{5/2}\kappa_C
+
(v_w-1)^3\kappa_D
},
\label{eq:app_kappa_sw}
\end{equation}
valid for $v_J\leq v_w$, with
\begin{align}
\kappa_C
&\simeq
\frac{\sqrt{\alpha}}{0.135+\sqrt{0.98+\alpha}},
\label{eq:app_kappa_C}
\\
\kappa_D
&\simeq
\frac{\alpha}{0.73+0.083\sqrt{\alpha}+\alpha},
\label{eq:app_kappa_D}
\\
v_J
&=
\frac{1}{1+\alpha}
\left(
\frac{1}{\sqrt{3}}
+
\sqrt{\alpha^2+\frac{2\alpha}{3}}
\right).
\label{eq:app_vJ}
\end{align}

For the benchmark value $v_w=1$, this interpolation reduces to
\(\kappa_{\rm sw}=\kappa_D\).

\paragraph{Turbulence.}
 The magnetohydrodynamic-turbulence contribution is evaluated as
\begin{equation}
\Omega_{\rm turb}h^2(f)
=
3.35\times 10^{-4}
\,v_w\,
\frac{H_n}{\beta}
\left(
\frac{\kappa_{\rm turb}\alpha}{1+\alpha}
\right)^{3/2}
\left(\frac{100}{g_*}\right)^{1/3}
\frac{(f/f_{\rm turb})^3}
{\left(1+f/f_{\rm turb}\right)^{11/3}
\left(1+8\pi f/h_*\right)}.
\label{eq:app_Omega_turb}
\end{equation}
The corresponding characteristic frequency and redshifted Hubble frequency are
\begin{align}
f_{\rm turb}
&=
27\,\mu{\rm Hz}\,
\frac{1}{v_w}
\frac{\beta}{H_n}
\left(\frac{T_n}{100\,{\rm GeV}}\right)
\left(\frac{g_*}{100}\right)^{1/6},
\label{eq:app_fturb}
\\
h_*
&=
16.5\,\mu{\rm Hz}\,
\left(\frac{T_n}{100\,{\rm GeV}}\right)
\left(\frac{g_*}{100}\right)^{1/6}.
\label{eq:app_hstar}
\end{align}
Following the prescription used in the main text, we take
\begin{equation}
\kappa_{\rm turb}=0.1\,\kappa_{\rm sw}.
\label{eq:app_kappa_turb}
\end{equation}

In the numerical analysis we set $v_w=1$, as stated in
Sec.~\ref{subsec:gw_spectrum}. Equations~\eqref{eq:app_Omega_sw} and
\eqref{eq:app_Omega_turb} are the spectral functions used to generate the
benchmark curves shown in Figs.~\ref{fig:gw_scenario_I} and
\ref{fig:gw_scenario_IV}.

\acknowledgments

B. L. Sánchez-Vega acknowledges financial support from the Minas Gerais State Research Support Foundation, FAPEMIG, under grant APQ-02289-26. I. M. Pereira and J. P. Carvalho-Corrêa thank the National Council for Scientific and Technological
Development of Brazil, CNPq, for financial support through Grants n°  161469/2021-3 and 141118/2022-9, respectively. A.
C. D. Viglioni and J.P. Cunha-Melo thank the Coordination for the Improvement of Higher
Education Personnel, CAPES, for financial support.

\bibliographystyle{JHEP}
\bibliography{refs.bib}

\end{document}